\documentclass{pasa}

\usepackage{graphicx}	

\def\arcsec   {\hbox{$^{\prime\prime}$}}
\def\degr     {\hbox{$^\circ$}}
\def\arcmin   {\hbox{$^{\prime}$}}
\newcommand{\fig}{Fig.}
\newcommand{\figs}{Figs.}
\newcommand{\tab}{Table}

\newcommand{\sect}{Section}
\newcommand{\sects}{Sections}
\newcommand{\chap}{Chapter}

\newcommand{\eqn}{Equation}

\newcommand{\Fig}{Fig.}

\newcommand{\Tab}{Table}

\newcommand{\paperone}{Paper~I}
\newcommand{\papertwo}{Paper~II}
\newcommand{\paperthree}{Paper~III}
\newcommand{\paperfour}{Paper~IV}
\newcommand{\deltane}{\delta N_e^2}

\title[IPS with the MWA V]{Interplanetary Scintillation with the Murchison Widefield Array V: An All-sky Survey of Compact Sources using a Modern Low-frequency Radio Telescope}

\author[J. S. Morgan et al.]{J. S. Morgan,$^{1}$ J-P. Macquart,$^{1}$ R. Chhetri,$^{1}$
R. D. Ekers,$^{1,2}$ S. J. Tingay$^{1}$ and E. M. Sadler$^{3}$ \\
\affil{$^{1}$International Centre for Radio Astronomy Research, Curtin University, GPO Box U1987, Perth, WA 6845, Australia}
\affil{$^{2}$CSIRO Astronomy and Space Science (CASS), P.O. Box 76, Epping, NSW 1710, Australia}
\affil{$^{3}$Sydney Institute for Astronomy, School of Physics A28, The University of Sydney, NSW 2006, Australia}
}

\jid{PASA}
\doi{10.1017/pas.\the\year.xxx}
\jyear{\the\year}

\usepackage{aas_macros}
\usepackage{hyperref} 
\hypersetup{colorlinks,citecolor=blue,linkcolor=blue,urlcolor=blue}

\hypersetup{draft}

\begin{document}

\begin{frontmatter}
\maketitle
\begin{abstract}
We describe the parameters of a low-frequency all-sky survey of compact radio sources using Interplanetary Scintillation (IPS), undertaken with the Murchison Widefield Array (MWA).
While this survey gives important complementary information to low-resolution survey such as the MWA GLEAM survey, providing information on the subarsecond structure of every source, a survey of this kind has not been attempted in the era of low-frequency imaging arrays such as the MWA and LOFAR.
Here we set out the capabilities of such a survey, describing the limitations imposed by the heliocentric observing geometry and by the instrument itself.
We demonstrate the potential for IPS measurements at any point on the celestial sphere and we show that at 160\,MHz, reasonable results can be obtained within 30\degr\ of the ecliptic (2$\pi$ str: half the sky).
We also suggest some observational strategies and describe the first such survey, the MWA Phase I IPS survey.
Finally we analyse the potential of the recently-upgraded MWA and discuss the potential of the SKA-low to use IPS to probe sub-mJy flux density levels at sub-arcsecond angular resolution.
\end{abstract}

\end{frontmatter}



\section{Introduction} \label{sec:intro}

Interplanetary Scintillation (IPS) was discovered in 1964 by \citet{Clarke:phdthesis} and was immediately identified by \citet{1964Natur.203.1214H} as an incredibly powerful astrophysical tool.
In clearly discriminating those sources with significant flux density coming from spatial scales $\lesssim$0.3\arcsec, the technique utilises the turbulent ionised interplanetary medium as an interferometer $\sim$1000\,km in extent.
This allowed small spatial scales to be probed at low-radio frequencies long before the first Very Long Baseline Interferometry (VLBI) observations.
For example, \citet{1965Natur.207...59H} used IPS to prove the existence a very compact source in the Crab Nebula, with a brightness temperature high enough to preclude synchrotron emission \citep[later identified as the Crab Pulsar:][]{1968Sci...162.1481S,1969Natur.221..453C}.

Even once VLBI was developed, IPS remained competitive for probing source sizes at low frequencies.
\citet{1972Natur.236..440R} used IPS measurements to measure scatter broadening due to the Interstellar Medium.
This work was continued by \citet{1984Natur.312..707R} to measure the much stronger scattering at very low Galactic latitudes (<5\degr), while \citet{1992Natur.355..232H} used IPS measurements over the whole sky to find early evidence for the ``Local Bubble'' \citep[e.g.][]{1998ApJ...500..262B}. 

Even today, when radio interferometry on much longer baselines is routine, this approach is still a useful one, since most of the technical challenges of long baseline interferometry are avoided: all that is required is the ability to measure source brightness at high (>1Hz) cadence.
In \citet{2018MNRAS.473.2965M} (hereafter \paperone) we outlined an approach to IPS studies that is optimised for low-frequency large-N imaging arrays.
We showed that with an instrument such as the Murchison Widefield Array (MWA), with extremely wide fields of view (FoV) and excellent instantaneous imaging fidelity, we can measure the IPS signature, and hence quantify subarcsecond structure, for many thousands of sources simultaneously.
\citet{2018MNRAS.474.4937C} (hereafter \papertwo) have applied this technique to a field with full overlap with the GLEAM survey \citep{2015PASA...32...25W,2017MNRAS.464.1146H}.
They show that IPS measurements in combination with the spectral information provided by a broadband continuum survey are a powerful combination for understanding the nature of a wide variety of astrophysical sources.
\citet{2018MNRAS.479.2318C} (hereafter \paperthree) have developed a technique for measuring the source counts of compact sources directly from IPS data.
This analysis shows the extent to which peaked and Compact Steep Spectrum (CSS) sources dominate the compact source population at low frequencies.
Sadler et al. (submitted, MNRAS, hereafter \paperfour) show that the strong scintillators detected by Chhetri et al. in \papertwo\ have a relatively high median redshift ($z\sim$ 1.7) and suggest that the IPS technique may be a powerful way to detect High-z Radio Galaxies.

These pilot studies with the MWA have important implications for the future of low-frequency radio astronomy, since they suggest a low-cost method to open up a powerful capability for sub-arcsecond measurements of radio sources for instruments with much lower interferometric resolution.
However, the four previous papers in this series have been based on just two 5-minute IPS observations.
A survey of the full sky on a regular (daily to weekly) cadence has the following advantages:
1) by observing each source multiple times, stochastic variations in the solar wind are averaged over, reducing the uncertainty in estimating the compact flux density; 
2) sources are observed over a range of solar elongations, increasing sensitivity and providing an additional parameter which may be used to calculate source size; and
3) the sky is well sampled, providing a far wider survey area and a much more uniform sensitivity than is possible with a single pointing.
Additionally, regular observations of IPS from a given set of sources provide a wealth of information on the heliosphere, however this is outside the scope of this work.

In this paper, we aim to set out the capabilities of an \emph{astrophysical} IPS survey with a modern low-frequency instrument.
In \sect~\ref{sec:ips} we draw on the results of many decades of IPS observations (and in-situ measurements) of the heliosphere to provide a description of the solar wind which is sufficiently detailed to allow IPS observations to be properly interpreted; we consider the geometry of IPS observations from the Earth over the course of a full year and solar cycle, and describe the IPS observations that can be made for each point of the celestial sphere.
In \sect~\ref{sec:mwa} we describe the instrumental limits that the MWA imposes on our observations, in particular primary beam considerations and how to optimally make observations while dealing with the presence of the Sun.
In \sect~\ref{sec:survey} we describe the MWA IPS observing campaign carried out from December 2015 to July 2016, covering RA 15h--24h; 0h--8h, revisiting fields with cadences between one hour and one week.
Finally in \sect~\ref{sec:future} we consider the prospects for future IPS surveys, in particular the utility of IPS observations for astrophysical study in the era of the Square Kilometre Array (SKA).

\section{The Solar Wind and the observational Geometry of IPS} \label{sec:ips}
In this section we will draw on the literature to provide a detailed 3D picture of the solar wind as it changes with time (through both stochastic variations in space weather and over the solar cycle).
This provides the background necessary to allow us to predict sensitivity and other critical parameters of an IPS survey (see \sect~\ref{sec:survey}).
\subsection{The Solar Wind} \label{sec:solarwind}
The solar wind is a supersonic outflow of plasma from the Sun, travelling at speeds $\sim$ 400--800\,km/s.
The rotation of the Sun means that solar wind structures emanating from a particular point on the Sun's surface (and the magnetic field lines) describe an Archimedian spiral known as the Parker Spiral \citep{1958ApJ...128..664P}.
The spiral crosses the Earth's orbit at an angle of approximately 45\degr East of the Sun; however as viewed from the Earth, a solar wind feature travelling from the Sun to Earth barely moves on the celestial sphere, since the Earth's orbital speed is almost negligible compared to typical solar wind speeds.

The density of the radial solar wind would be expected to be proportional to $R^{-2}$ (where $R$ is distance from the Sun).
The relevant density for determining how electromagnetic waves propagate through the plasma is the electron number density.
The variance of the electron density fluctations $\deltane$ as a function of distance from the Sun would then be
\begin{equation}
	\deltane \propto R^{-\beta}
	\label{eqn:beta}
\end{equation}
 with an exponent, $\beta$, of $4$ \citep{Thompson:2001}.
\citet{1971MNRAS.155..185R} give a power law index of $4.1\pm0.1$ based on IPS observations.
Observations at 326\,MHz by \citet{1993SoPh..148..153M} over a range of conditions agree with this value within errors.
Much closer to the Sun (2.7\degr--13.3\degr), \citet{1995ApJ...445..999S} derived a value of  $\beta=3.7\pm0.3$ from VLBI observations of phase scintillations.

The IPS phenomenon itself arises from density fluctations on scales $\sim$10--1000\,km.
These fluctuations have been studied in detail using a variety of remote sensing techniques, as well as in-situ measurements \citep{1978SSRv...21..411C}, especially close to the Sun \citep{1989ApJ...337.1023C} where the turbulence is approximately described by a power law, steepening towards smaller scales.
This steepening is also seen at longer elongations \citep{1994JGR....9923411M}.
Beyond 1\,AU the turbulence continues to evolve towards a Kolmogorov spectrum \citep{2010JGRA..11512101R}

The most obvious anisotropy in the solar wind is the existence of the equatorial and polar streams of the solar wind.
The polar wind is perhaps 50\% faster \citep{1980Natur.286..239C} and 10--15$\times$ less dense than the equatorial solar wind \citep{1995JGR...10017069C}.
At solar maximum the volume that the polar wind fills shrinks to just a few percent.
However in the declining phase and at solar minimum the polar solar wind is more dominant.
The spectrum of density fluctuations is somewhat different for high- and low-speed streams and varies depending on solar activity and over the solar cycle \citep{1993SoPh..148..153M,1994JGR....9923411M}.

In addition to these large-scale structures, there are also transient changes in the solar wind, the most notable of which are Coronal Mass Ejections (CMEs).
IPS observations also reveal stochastic changes in scintillation index, implying variations in the mean plasma density.
We loosely refer to all of these departures from an isotropic solar wind as ``space weather''.

\subsection{Scintillation} \label{sec:scintillation}
When electromagnetic waves cross an inhomogeneous ionised medium, local density fluctuations will cause the radiation to be deflected.
At a sufficient distance, this will cause increases and decreases in the observed flux density, as radiation that has taken slightly different paths to the observer interferes.
We give a brief description relevant to IPS here and refer the reader to \citet{1993ppl..conf..151N} for further details (see also \citealp{Thompson:2001} \chap~13 and the discussion in \sect~3.4 of \papertwo).

If the scattering is ``weak'', only radiation crossing the scattering medium within a radius $r_F$ of the point on the Line of Sight (LoS) from source to observer can contribute to scintillation (where $r_F$ is the Fresnel Scale).
The weak regime applies when $r_{\mathit{diff}}\gg r_F$ where $r_{\mathit{diff}}$ is the transverse distance over which the scattering medium produces an rms phase change of one radian.
Scintillation in the weak regime is relatively simple to model, even when the scattering occurs at many locations along the line of sight (LoS), thus most IPS observations are undertaken in the weak regime.
The strength of scintillation can be described by the scintillation index $m$, which is the scintillating flux density (standard deviation) as a fraction of the mean flux density.
In weak scintillation $m\ll1$, and $m$ approaches 1 as $r_{\mathit{diff}}$ approaches $r_F$.
If the source of radiation is in the far field, then $\theta_F$, the angle subtended by $r_F$ at the location of the observer, is given by
\begin{equation}
	\theta_F = \sqrt{\frac{\lambda}{2\pi d}}
	\label{thetaf}
\end{equation}
where $\lambda$ is the observed wavelength and $d$ is the distance from the scattering medium to the observer.
This gives a value $\sim$ 0.3\arcsec\ for metre-wavelength IPS observations.
Sources with a size $\theta_s$ < $\theta_F$ will exhibit scintillation like a point source.
Larger sources have lower scintillation indices than a point source.

In the strong scintillation regime ($r_{\mathit{diff}}\ll r_F$), a sufficiently compact source will show diffractive scintillation with a scintillation index of approximately unity\footnote{In theory, modulation indices can exceed 1 by 5\% but this is rarely, if ever observed \citep{1978SSRv...21..411C}.
While we have observed scintillation indices >1 with the MWA this is probably attributable to errors in the measurement of the scintillating or mean flux density.}.
However, the critical angular size for a source to be compact is now $\theta_{\mathit{diff}}$ rather than $\theta_F$.
Another characteristic of diffractive scintillation is its narrow fractional bandwidth $\delta\nu/\nu\sim\left(r_{\mathit{diff}}/r_F\right)^2$.

\subsection{IPS Geometry} \label{sec:geometry}
\begin{figure}
  \includegraphics[width=\columnwidth]{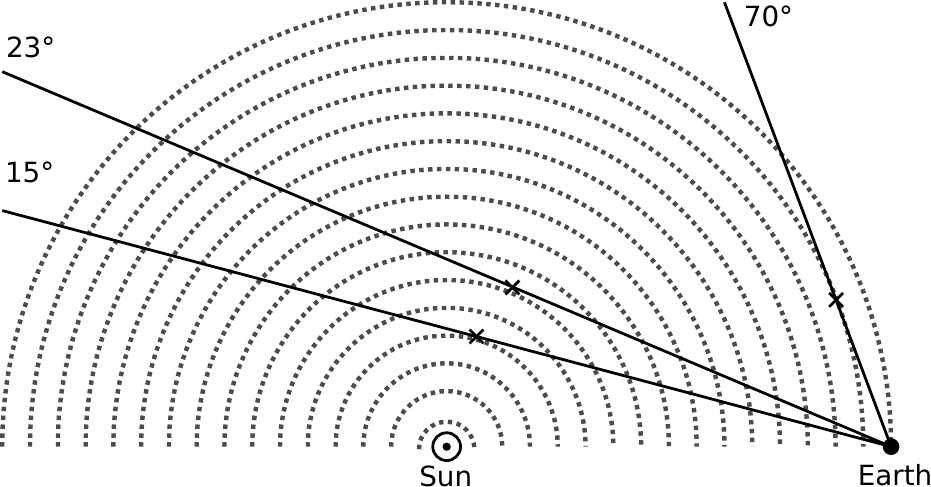}\,
  \caption{Following \citet{Bell:phdthesis}, the geometry of IPS. Dotted lines show solar radii from 1/16 AU to 1 AU. Solid lines show the lines of sight for transition into the strong regime at 80\,MHz (23\degr) and 162\,MHz (15\degr) and the approximate longest elongation with well-defined piercepoint (70\degr). Crosses mark the point of closest approach to the Sun}
  \label{fig:ips_geometry}
\end{figure}
\fig~\ref{fig:ips_geometry} illustrates the basic geometry of IPS close to the Sun (elongation, $\epsilon\lesssim70$\degr).
The LoS from the source to the Earth passes closer than 1\,AU to the Sun.
In principle, solar wind turbulence all the way along the LoS contributes to the IPS signal.
However IPS can, to first order, be considered to come from the point of the closest approach of the LoS to the Sun, due the fact that the turbulence of the solar wind is so much stronger close to the Sun \citep{1966MNRAS.134..221L}.
The distance of closest approach to the Sun is the impact parameter $p$:
\begin{equation}
	p=\sin\left(\epsilon\right)\textrm{\,AU},
	\label{eqn:p}
\end{equation}
whereas the distance from Earth to the screen $d$ is:
\begin{equation}
	d=\cos\left(\epsilon\right)\textrm{\,AU}.
	\label{eqn:d}
\end{equation}

\begin{figure}
  \includegraphics[width=\columnwidth]{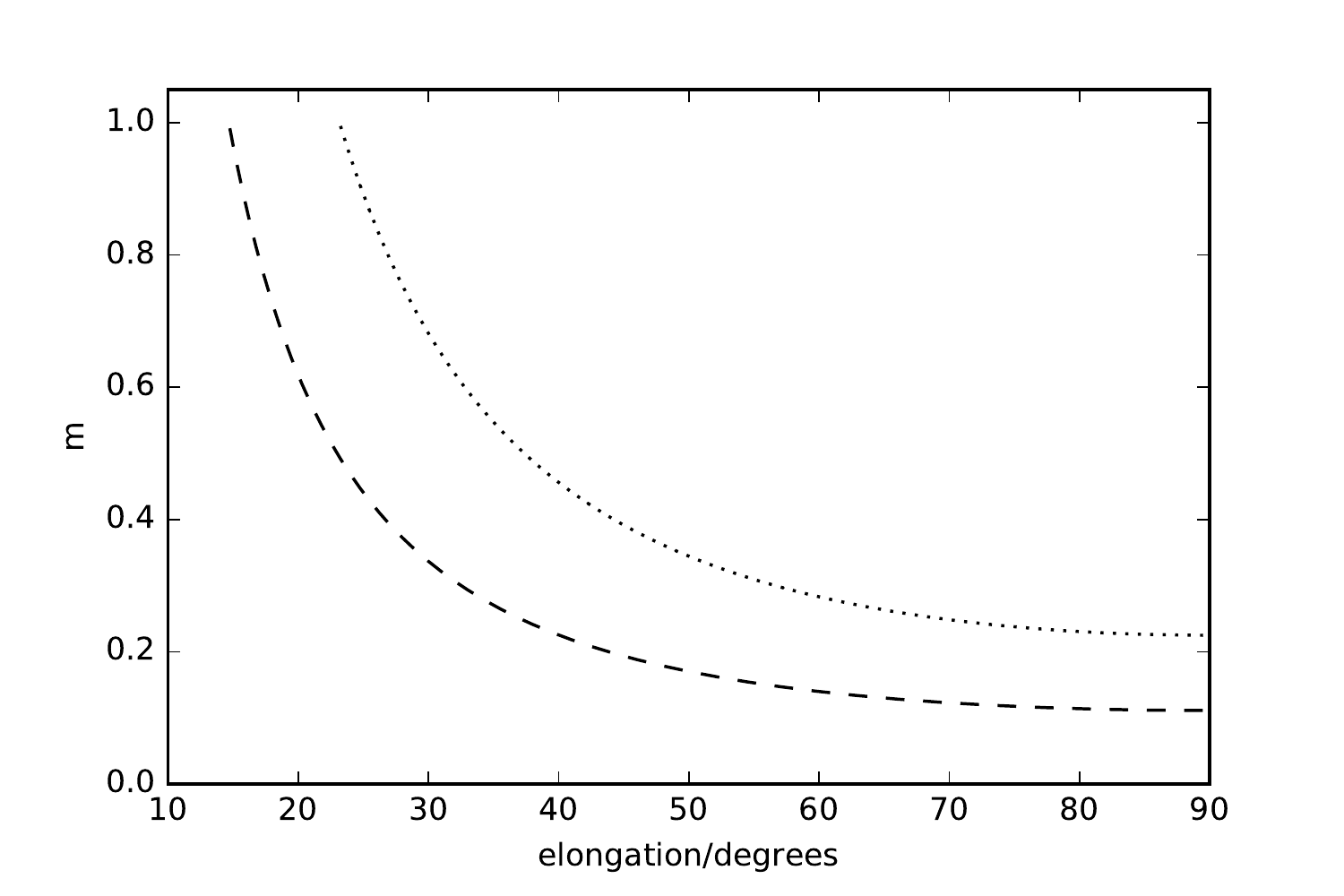}\,
  \caption{Scintillation index $m$ as a function of solar elongation while scintillation remains in weak regime. Dashed line is for 162\,MHz, dotted line is for 80\,MHz.} 
  \label{fig:m_p}
\end{figure}

An empirical relationship between $p$ and the average strength of scintillation was measured by \citet{1969P&SS...17..313H} and has been shown to be valid over a wide range of frequencies and solar elongations \citep{1973JGR....78.1543R}.
For a sufficiently compact source ($\lesssim$0.3\arcsec), the measured scintillation index $m$ is given by
\begin{equation}
	m=0.06\lambda^{1.0}p^{-b},
	\label{eqn:rickett}
\end{equation}
where $\lambda$ is the wavelength in m, and $b\approx1.6$.
This is shown in \fig~\ref{fig:m_p}.
With a radial solar wind as described by \eqn~\ref{eqn:beta}, $b=\left(\beta-1\right)/2$ \citep{1993SoPh..148..153M}.
This relationship only applies in the weak scintillation regime.
As a source approaches the Sun its scintillation index will increase until finally saturating and turning over as $m$ approaches 1.
Thus for a given frequency there is a range of solar elongations which are optimum for IPS observations: too far away and $m$ becomes too small to be easily measurable; too close and the strong scintillation regime is reached.
High radio frequencies can be used to probe very close to the Sun \citep[e.g.][]{1995sowi.conf...62T}, while at the lower frequencies, IPS can be observed all over the sky \citep[e.g.][]{Bell:phdthesis}.

The preceding description is only valid for day-side IPS observations close to the Sun.
At longer elongations ($\epsilon\gtrsim70$\degr) the distance from the Sun along the LoS does not vary as rapidly, and so the weighting function is not so concentrated on the piercepoint (see \fig~\ref{fig:ips_geometry}).
A fixed distance of 0.3\,AU may be a better estimate for the distance to the scattering screen \citep{1991JGR....96.1717R}.
IPS can also be observed on the nightside \citep[e.g.][]{1970MNRAS.150..141H,2015ApJ...809L..12K} however here the impact parameter is even less well-defined.

\subsection{IPS in the presence of space weather} \label{sec:ips2}
\eqn~\ref{eqn:rickett} assumes that the solar wind at a given distance from the Sun does not vary with time or solar latitude, which is not the case.
Taking data from two solar cycles, \citet[\fig~7]{1993SoPh..148..153M} find that at solar maximum, the solar wind is indeed radially symmetric on average, however at solar minimum the contour of average constant scattering describes an ellipsoid which is flattened at the poles.
Thus the mean scintillation index would be better described as
\begin{equation}
	m=0.06\lambda^{1.0}\left(e\,p\right)^{-b},
	\label{eqn:mano}
\end{equation}
where $e$ is the elliptical term and
\begin{equation}
	e=\sqrt{\rho^2\sin^2\phi + \cos^2\phi},
	\label{eqn:ellipse}
\end{equation}
where $\rho$ is the ratio of the equatorial to polar diameters of an ellipsoid of constant scattering (i.e. $\rho\ge1$: 1 at solar maximum, 1.5 at solar minimum) and $\phi$ is the latitude of the point on the surface of the Sun directly below the scattering screen.

We reiterate that this elliptical model represents a long-term average.
The polar solar winds giving rise to a decrease in scintillation towards the poles occupy different volumes of the heliosphere to the ecliptic solar winds \citep{1996Ap&SS.243...87C}.
Even when both streams are contributing to an IPS signal, multi-station IPS observations resolve two distinct peaks in the cross-correlation function, \citep{1996Natur.379..429G}.
Since the sector of the LoS responsible for IPS is relatively short, especially close to the Sun, the observed scintillation indices close to the boundary between polar and equatorial streams are likely to be quite bimodal, depending on whether the equatorial stream occupies the piercepoint or not (though where both are close to the piercepoint, the much denser equatorial stream will dominate the IPS signal).

In addition to the polar and equatorial solar winds, stochastic changes in solar wind turbulence also cause departures of the observed scintillating index $m_{obs}$ from the expected scintillation index $m$ (e.g. that predicted by \eqn~\ref{eqn:rickett}).
These variations are parametrised as the ``scintillation enhancement factor\footnote{See \citet{1986P&SS...34...93T} for further details, including the relationship between $g$ and the overall plasma density, determined via comparison of IPS observations and in situ measurements. See also the discussion \sect~3.3 of \citet{1993SoPh..148..153M} and references therein.}'' $g$, where 
\begin{equation}
	g = \frac{m_{obs}}{m} .
	\label{eqn:g}
\end{equation}
\citet{Hajivassiliou:phdthesis} \citep[as cited in][]{1990MNRAS.247..491H} characterised the distribution of $g$ values for IPS observations, decomposing them into more extreme events which are easily identified, and less extreme variations which are Gaussian distributed with a standard deviation of approximately 20\% at an elongation of 40\degr\ from the Sun.
This is consistent with the findings presented in \papertwo\ where it was shown that the scintillation indices of flat spectrum sources (which are expected to be compact) followed the relation given in \eqn~\ref{eqn:rickett} with a standard deviation of approximately 20\%. 

\subsection{Optimal detection of compact sources} \label{sec:optimal_detection}
We now briefly describe our method for detecting compact sources with IPS via interferometric imaging.
This is described in detail in \paperone, and the reader is referred to \sect~2.3 of that paper for further details.
Essentially, the full FoV of the instrument is imaged on a time cadence sufficiently short to critically sample IPS variability.
This results in an image cube with time as the 3rd axis.
The timeseries corresponding to each spatial pixel is then filtered to emphasise the scintillation timescale of IPS, downweighting variability more rapid than ~1\,s and slower than ~10\,s (the latter to filter out ionospheric variability).
The rms of each filtered timeseries is then calculated and an image is constructed from these values.
This ``variability image'' has Gaussian noise statistics, and sources detected at a particular level of confidence (e.g. 5-$\sigma$) can readily be extracted using standard techniques \citep[see e.g][]{2012MNRAS.422.1812H}.

A particular characteristic of searching for signals by their standard deviation alone identified in \paperone\ is that increased observing time will not lead to a great increase in sensitivity since sensitivity scales with observing time $T$ only as $T^{1/4}$, rather than the usual $\sqrt{T}$.
Since greater sensitivity cannot easily be attained by integrating for long, it becomes even more important to observe sources when their scintillation index is maximised.
In the remainder of this section we discuss the optimum solar elongation at which sources should be observed.

As discussed above, at a given observing frequency, the scintillation index of a sufficiently compact source will increase as the source approaches the Sun, up to the point where the strong scattering regime is reached.
Given that a sufficiently compact source will show diffractive scintillation with a scintillation index of unity, this might suggest that searching for sources in the strong scintillation regime will yield the strongest detections, however there are a number of reasons why this is not the case.
First, as noted in \sect~\ref{sec:scintillation}, the critical size for sources to show scintillation is smaller for diffractive scintillation than for weak.
If particularly compact sources (e.g. pulsars) are being searched for, this may be a useful extra constraint.
However, in general fewer sources will be detectable.

The narrow bandwidth of diffractive scintillation also introduces complications \citep[see][for a dynamic spectrum of IPS partly in the strong scintillation regime]{2013SoPh..285..127F}. 
If the observing bandwidth exceeds the scintillation bandwidth then it will be necessary to form multiple images across the band to avoid averaging out the scintillations.
This is technically feasible for modern correlators, however it will incur extra processing costs.
Additionally, when searching for the IPS signal via its standard deviation, searching over $N$ spectral images reduces sensitivity by $N^{1/4}$, so this should be avoided.

Finally, in the diffractive scintillation regime, the timescale of scintillation will reduce.
Our 0.5\,s sampling time is only just sufficient to critically sample IPS in the weak regime (the IPS signal remains strong up to at least 0.3\,Hz; see \sect~2.2.2 of \paperone) .
Thus faster sampling would be required to fully capture the IPS signature in the strong regime, \emph{and} this would further reduce sensitivity due to the $T^{1/4}$ problem.

Modeling the scintillation in the transition from weak to strong scattering is difficult, although it has been calculated for the case of thin-screen, isotropic, Kolmogorov turbulence by \citet{2006ApJ...636..510G}.
For the remainder of this work we take the maximum scintillation index of typical IPS to be 80\% as measured empirically in \paperone, and we assume that this value is attained at the solar elongation predicted by \eqn~\ref{eqn:mano} (i.e. 27\degr\ and 18\degr\ at 80\,MHz and 162\,MHz respectively for a source on the ecliptic).

\begin{figure*}
  \includegraphics[width=\textwidth]{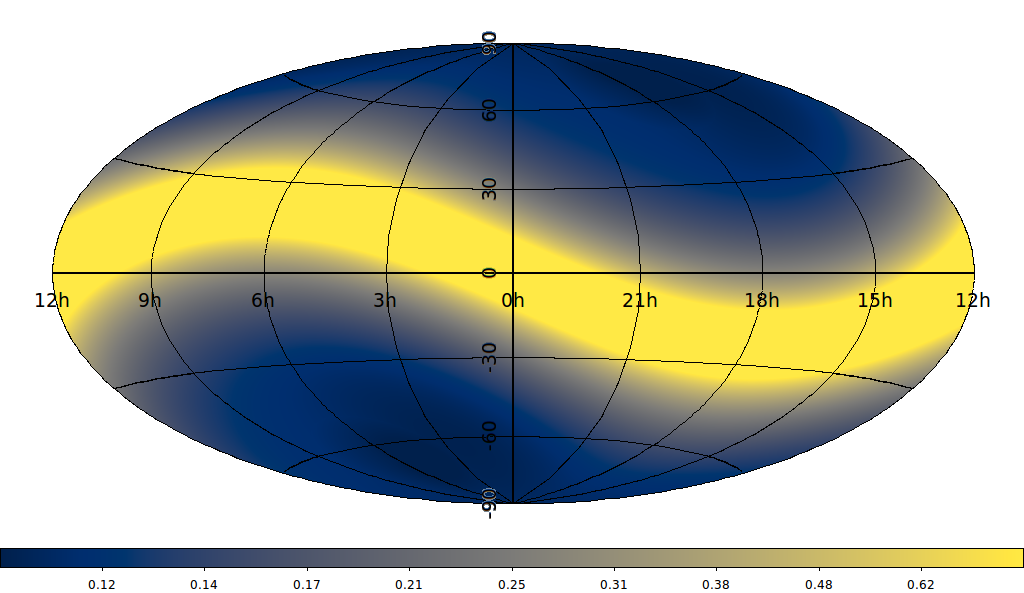}\,
  \includegraphics[width=\textwidth]{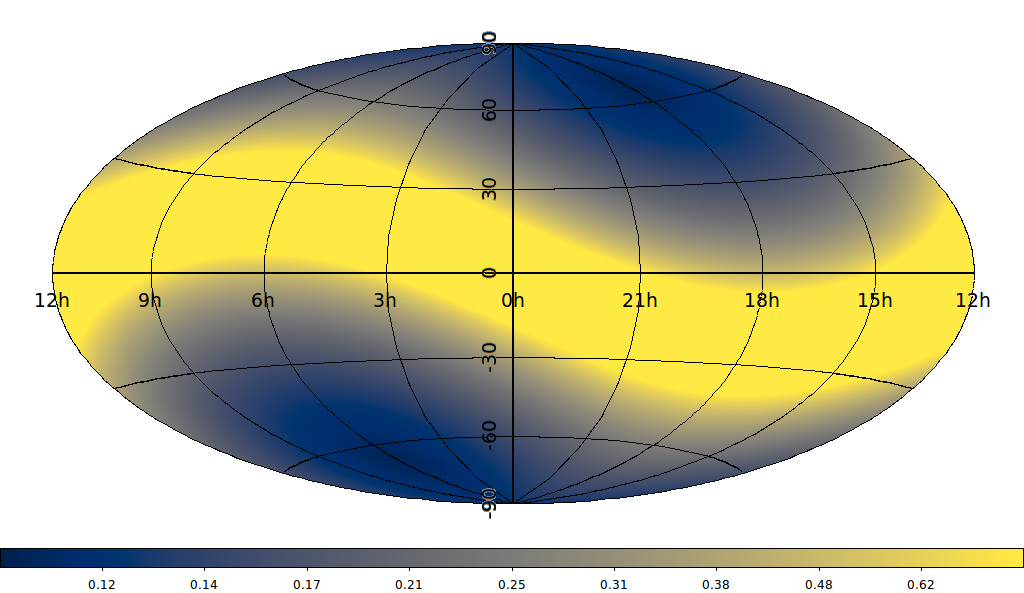}
  \caption{Maximum point-source scintillation index obtained during the year for every point on the celestial sphere for observations at 162\,MHz (See \eqn~\ref{eqn:mano}). Equatorial Coordinates (RA in hours, Declination in degrees). Inverse Hyperbolic sine scale. Top panel is for solar minimum ($\rho=1.5$), bottom panel is for solar maximum ($\rho=1.0$).}
  \label{fig:m_hi}
\end{figure*}
\begin{figure*}
  \includegraphics[width=\textwidth]{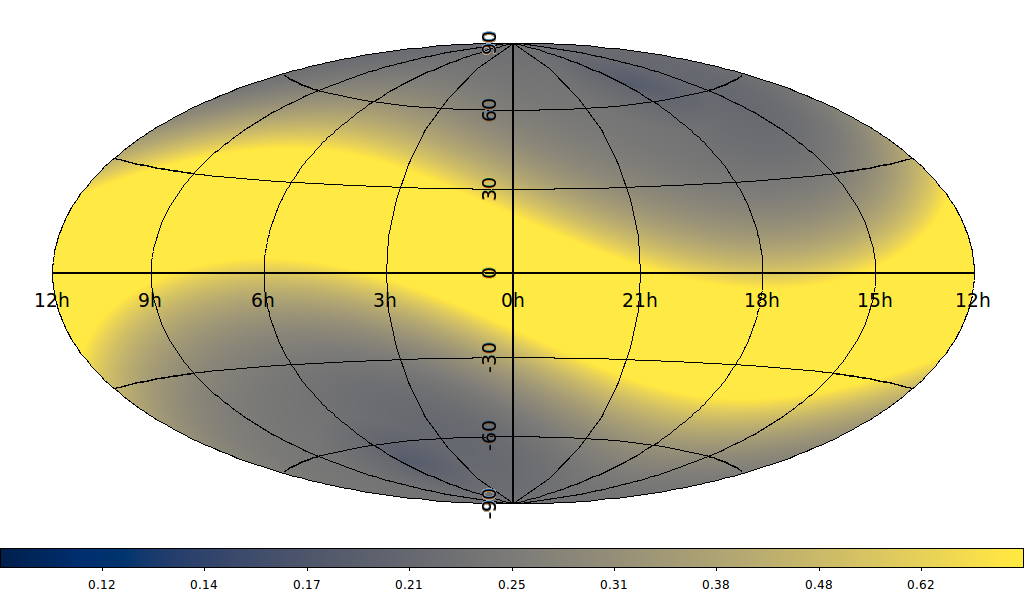}\,
  \includegraphics[width=\textwidth]{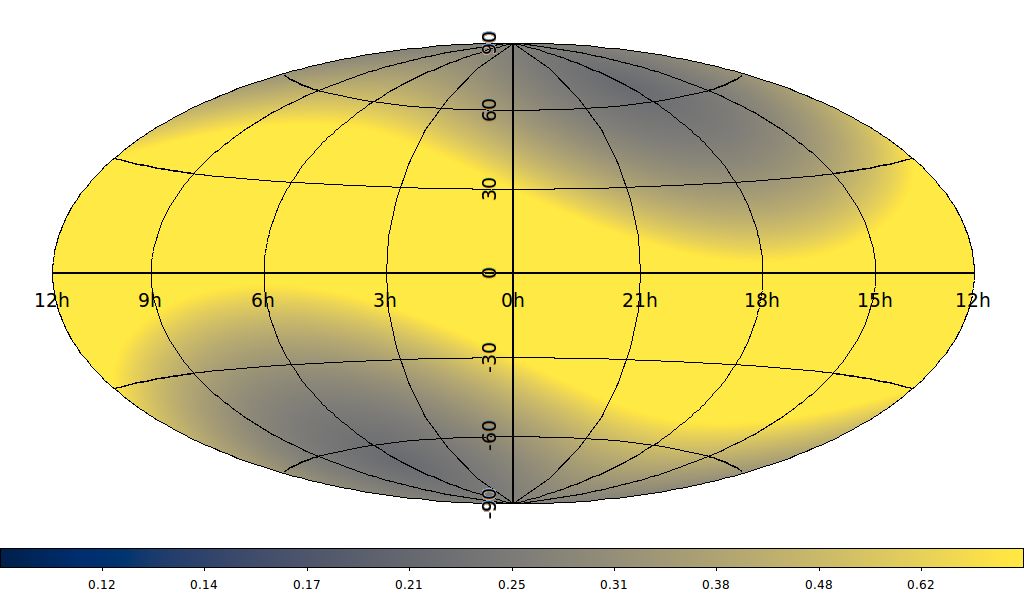}
  \caption{Maximum point-source scintillation index obtained during the year for every point on the celestial sphere for observations at 80\,MHz (See \eqn~\ref{eqn:mano}). Equatorial Coordinates (RA in hours, Declination in degrees). Inverse Hyperbolic sine scale. Top panel is for solar minimum ($\rho=1.5$), bottom panel is for solar maximum ($\rho=1.0$).}
  \label{fig:m_lo}
\end{figure*}
In \figs~\ref{fig:m_hi} and \ref{fig:m_lo}, we combine these limits with \eqn~\ref{eqn:mano} to show the maximum scintillation index that is reached over the whole year (i.e. the scintillation index at the point of closest approach to the Sun) for a point source for any location on the celestial sphere, both at solar maximum and at solar minimum, for frequencies of 80 and 162\,MHz\footnote{For the purposes of calculating the solar latitude of the piercepoint, the North pole of the Sun points towards approximately RA 19h and Decl. +64\degr\ \citep{2002CeMDA..82...83S}.
This is approximately 7\degr\ from the ecliptic North Pole.
See \citet{2006A&A...449..791T} for a comprehensive treatment of solar coordinate systems.}.

The preceding discussion assumes that we are able to observe a source at whatever solar elongation we wish.
This is only the case for sources on the ecliptic.
Sources at higher ecliptic latitudes may not approach the Sun closely enough to reach the strong regime, particularly at solar minimum.
Furthermore their solar elongation will change relatively slowly over the year compared to a source on the ecliptic.
If such regions are surveyed a sufficient number of times, many of the detections will be due to random increases in $\delta N_e$ causing sources that would usually be below the detection limit to be upscattered.
A $g$-levels of 2.5 (which could increase the scintillation index of a source from 40\% to 100\%) is possible (see Paper I and \citealp{1986P&SS...34...93T}).
However, whether or not such occasional detections can lead to an estimate of the compact flux density of the source will depend on the extent to which the scattering along the LoS can be characterised.

\subsection{Determining the structure of compact sources} \label{sec:source_size}
In this section we discuss to what extent the structure of a compact source can be determined from IPS observations.
In \papertwo\ the only IPS observable considered was the scintillation index, $m$.
We then used \eqn~\ref{eqn:rickett} to compare the scintillation index of each source to that which would be expected from a point source, and termed this the Normalized Scintillation Index (NSI); thus a source with an NSI of 1 would be compact, whereas a source with NSI of 0.5 would have half its flux density coming from scales smaller than $\theta_F$.
Given that variations in the solar wind imposed relatively large errors ($\approx20\%$) on the NSI, we classified all of our sources into 3 broad categories: strong scintillators (NSI$\ge0.9$), which were dominated by a compact source, moderate scintillators ($0.4\le$NSI$<0.9$) and sources with weak or no scintillation (NSI$<0.4$).

With multiple observations of each source, it is possible to determine the NSI more accurately.
However there are still a multitude of possible source structures consistent with any given NSI.
\begin{figure}
  \includegraphics[width=\columnwidth]{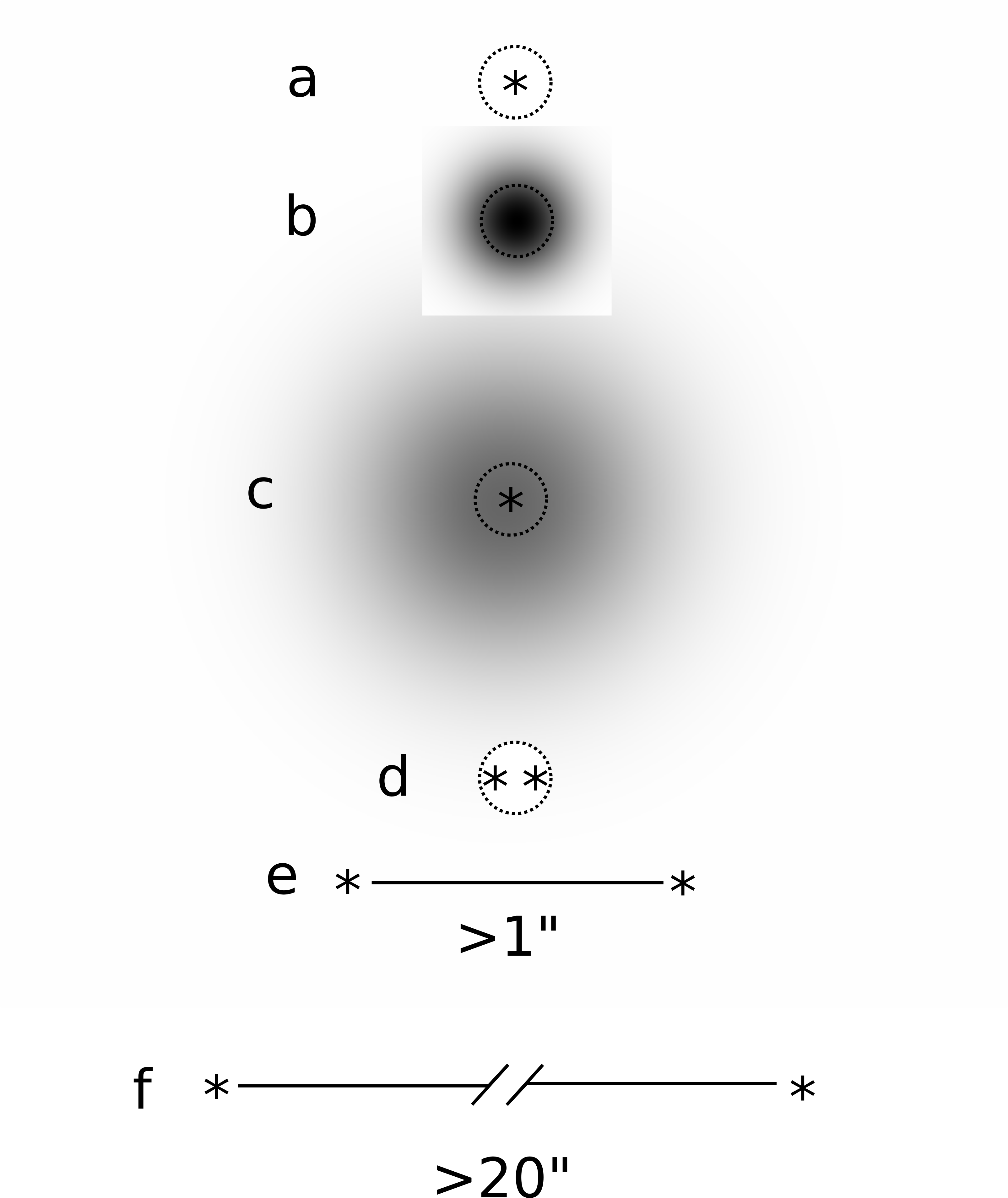}\,
  \caption{Illustrative source types. Asterisks indicate point-like source, greyscale represents extended emission, dotted circles indicate Fresnel scale ($\approx 0.3\arcsec$). Source types are: (a) ``point source'', (b) ``slightly extended source'', (c) ``embedded source'', (d) ``unresolved double'', (e) ``resolved double'', (f) ``wide double''}
  \label{fig:sources}
\end{figure}
\begin{table*}
  \footnotesize
  \centering
  \caption{\label{tab:source_nsi} NSI for source types illustrated in \fig~\ref{fig:sources}}
  \begin{tabular}{lrl}
    \hline
    Source type       & NSI                                   & Explanation \\
    \hline                                                                                                                           
    Unresolved        &  1                                                                              & -- \\
    Slightly Extended  &  $\theta_F/\theta_s$                                                            & Ratio of Fresnel diameter to source diameter \\
    Embedded          &  $S_c/\left(S_c+S_e\right)$                                                     & Ratio of compact (c) flux to total flux \\
    Unresolved double &  $\approx$1                                                                     & Note that this applies to \emph{any} structure <$\theta_F$ \\
    Resolved double   &  0.7--1                                                                         & Depends on separation and possibly orientation w.r.t solar wind velocity \\
    Wide double       &  $\sqrt{S^2_1+S^2_2}/\left(S_1+S_2\right)$                                      & >0.7. Can easily be extended to multiple components \\
    \hline
  \end{tabular}
\end{table*}
Some illustrative examples are shown in \fig~\ref{fig:sources} and the indicative NSI for each is given in \tab~\ref{tab:source_nsi}.
We define them as follows: a ``point source'' is a source where all emission comes from scales much smaller than the Fresnel Scale.
A ``slightly extended source'' is one where the emission comes from a scale comparable with the Fresnel scale.
An ``embedded source'' is a point source embedded in much more extended structure.
An ``unresolved double'' is a double source where the double separation is smaller than the Fresnel scale.
A ``resolved double'' is a double source where the double separation is larger than the Fresnel scale.
A ``wide double'' is a double source sufficiently large that its structure can be inferred by other means; for example The GMRT Sky Survey Alternative Data Release \citep{2017A&A...598A..78I}.

Real radio sources that could be approximated by each of these examples can be found, so none can be ruled out a priori.
\citet{2016A&A...595A..86J} used LOFAR \citep{2013A&A...556A...2V} at $\sim$150\,MHz to search for compact calibrators and found a noticeable reduction in correlated flux density on baselines corresponding to spatial scales $\sim$0.7\arcsec--2\arcsec.
This would suggest that at least some compact sources are slightly extended.

The NSI is not the only observable that an IPS survey would provide.
More generally the IPS signal itself is convolved with the source brightness distribution on arcsecond scales, so in principle information on more complex morphologies may be gleaned from IPS.
Below we consider various other observables and how they can be used to determine source structure.

\subsubsection{Calculating source size using m-p curves}
Previous IPS surveys determining the size of the scintillating component (assuming a Gaussian source brightness distribution) used the shape of the curve of scintillation index $m$ as a function of piercepoint $p$ (see \fig~\ref{fig:m_p}) \citep{1971MNRAS.155..185R,1987MNRAS.229..589P}.

However subsequently it was determined that this may lead to biased results \citep{1990MNRAS.247..491H} due both to the instrumental noise and uncertainties introduced by stochastic variations in the heliosphere.
The authors suggest that these errors due to the heliosphere may be mitigated by characterising the interplanetary medium using neighbouring sources (which the MWA is uniquely well-placed to do, see \sect~\ref{sec:model_helio}).

Even so, the shape of the curves for different source sizes are quite similar if they are scaled to the same peak value.
This means that this technique is unlikely to be useful in distinguishing slightly extended sources and embedded sources.

\subsubsection{Multi-frequency techniques}
\citet{1975ApJ...201..238W} has shown that observations at multiple frequencies yield independent information about the solar wind turbulence.
\citet{1983A&A...123..191S} showed that the correlation coefficient of the scintillation patterns at two frequencies separated by a factor of two depends only on the source size and the power law index of the turbulence.
They demonstrated convincingly that the source size can be recovered reliably in this way with observations of bright sources at 270 and 470\,MHz at solar elongations of 15\degr--37\degr.
Thus, the combination of an accurate NSI and the correlation coefficient should disambiguate slightly extended and embedded sources, and determine the size of the compact component.
One disadvantage of this approach is that it requires the scintillation to be in the weak regime at both frequencies, which limits the sky area over which this can be used. 
This also means that at the higher frequency the source must be detected strongly even though the scintillation index will be <0.5.

\subsubsection{Power Spectrum Analysis}
Assuming a solar wind whose level of turbulence is a known function of location in the heliosphere, the power spectrum of a source showing weak scintillation depends on the visibility of the source (usually assumed to be Gaussian), the spectrum of the turbulence in the solar wind (characterised by the power law index), the axial ratio of the turbulence, and the velocity of the solar wind \citep[see][and references therein]{1990MNRAS.244..691M}.
In principle, therefore, it is possible to model the scintillation due to each point along the LoS, integrate to find the power spectrum as observed at the Earth, and recover information about the source visibility by iterative fitting of model power spectra to the data.

Unfortunately, even for the simple case of assuming a Gaussian source brightness distribution, these parameters are strongly correlated.
For example, if the IPS signal varies more slowly than expected, this could be due to a lower solar wind speed, or a resolved source (the IPS signature is convolved with source structure along the solar wind velocity direction).
Furthermore, any departure of the true solar wind from the heliospheric model will introduce further uncertainties. 
Thus, the success of this approach will depend on how well the solar wind can characterised a priori.

\subsubsection{Distribution of fluctuations}
It is notable that pulsars (the most compact sources we can detect) show very strong signatures in skew and kurtosis (see \paperone, \fig~14).
For a slightly extended source, convolution of the IPS signature with the Gaussian source brightness distribution will blunt these higher-order moments.
Thus, this may be another method for distinguishing between slightly extended and embedded sources.

\subsubsection{Characterisation of source morphologies}
\begin{figure}
  \includegraphics[width=\columnwidth]{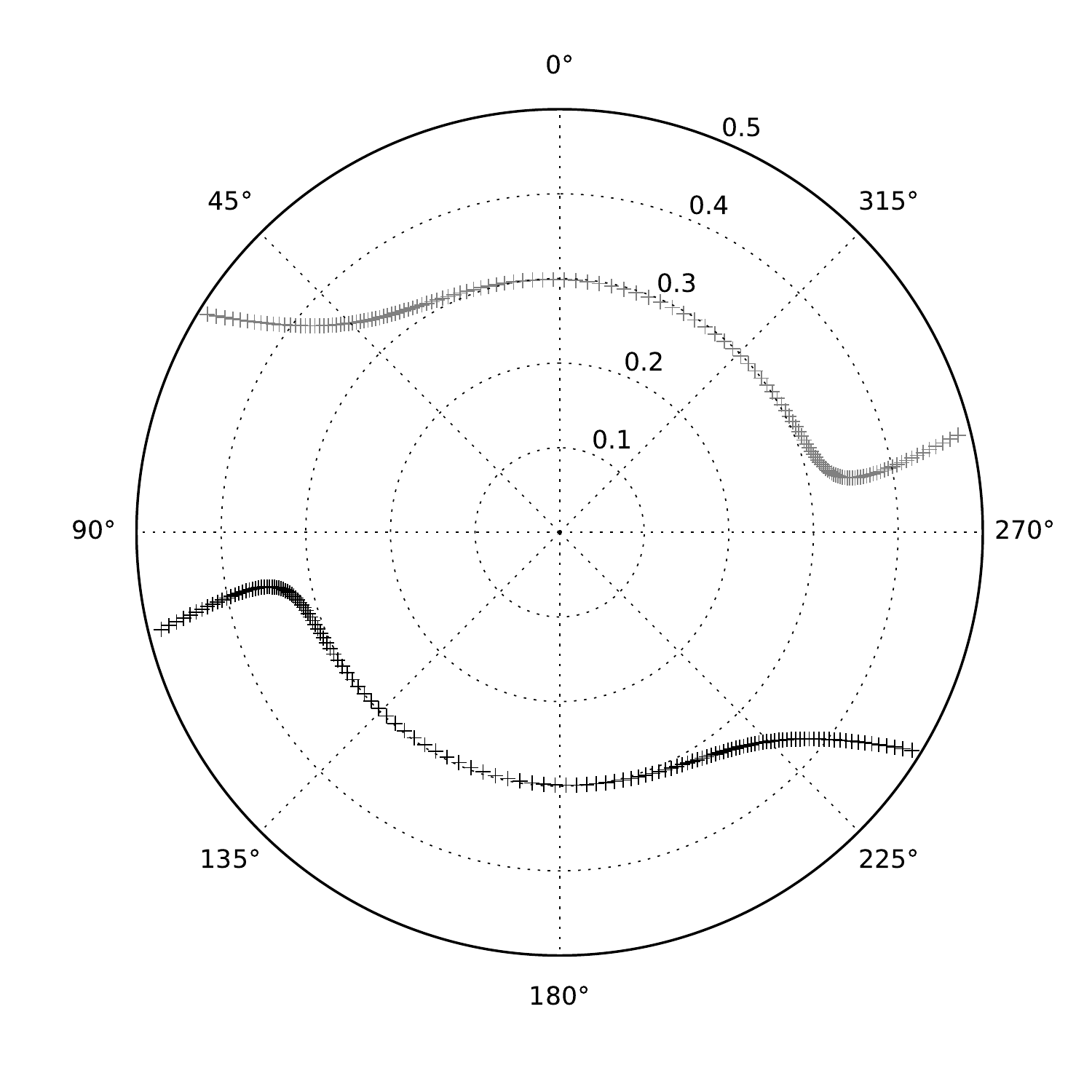}\,
  \caption{Polar plot of solar wind vector for a source located at RA 0, Decl. -20\degr (18\degr from the ecliptic). Radial distance shows $\theta_F$ in arcseconds. Each Black '+' indicates a daily observation while solar elongation <70\degr. The complex conjugate is plotted in grey to since these measurements would give almost identical information. This emphasises that $\sim$3/4 (rather than $\sim$3/8) of position angles are probed.}
  \label{fig:solar_vector_polar}
\end{figure}

In \paperone\ it was shown that a double structure on scales of a few arcseconds could be detected as a peak in the autocorrelation function \citep[c.f. detection of an emerging double via interstellar scintillation:][]{2007MNRAS.380L..20M}.
The angular separation of source components can be determined if the solar wind speed is known.
For sources off the ecliptic, the angle of the solar wind across the source will change with elongation. 
\Fig~\ref{fig:solar_vector_polar} shows the range of solar wind vectors encountered by a source 18\degr\ from the ecliptic while its elongation<70\degr.
Most directions across the source are probed, showing that for most double sources it will be possible to determine their major and minor axes and their position angle.
In some cases the solar wind has different turbulence in directions parallel and perpendicular to its velocity.
In these cases it may be possible to determine whether the sources are extended parallel or perpendicular to the solar wind velocity.
Therefore in principle it may be possible to reconstruct quite complicated sky brightness distributions.

\subsubsection{Modeling space weather} \label{sec:model_helio}
As discussed in \sect~\ref{sec:ips2}, the presence of the polar and equatorial streams, as well as stochastic variations in the heliosphere may complicate the astrophysical interpretation of IPS observations.
This potential difficulty in using IPS for astrophysical studies was identified early by \citet{Bell:phdthesis}.
For the simple task of identifying compact sources and determining their NSI, simply averaging over space weather with multiple observations is likely to be adequate.
However, with the possible exception of multi-frequency IPS, all of the techniques outlined above for more sophisticated characterisation of source structure are more critically affected by space weather.
This motivates us to consider to what extent the turbulence in the heliosphere can be modeled in order to deal with these nuisance parameters and recover the properties of the astrophysical sources more accurately.
A full exploration of these issues is beyond the scope of this paper, however below we give a brief overview of some possible approaches.

Since the solar wind emanates from the Sun, observations of the Sun at the time the region of the heliosphere of interest left the Sun can shed light on its properties.
Strong indication of which points along the LoS are occupied by the polar wind can be gleaned from white-light coronagraph images \citep{1996Ap&SS.243...87C,2008AnGeo..26.2229F}.
Reconstructions of the solar surface from IPS observations sometimes show astonishingly good correlation with observations of the Sun \citep[e.g.][]{2013PJAB...89...67T}, to the extent that it may be possible to use solar observations (such as magnetographs) to construct a 3D model of the heliosphere.

IPS observations are made routinely by a number of observatories around the world.
ISEE \citep{2011RaSc...46.0F02T} in particular also operates as a multi-station array for much of the year, and provides timely information on $g$-levels and solar wind velocities.
Where these observations are close enough in space and time, they could be used to make corrections.
The Murchison Radio Observatory (home of the MWA and planned SKA-low) is particularly well placed in longitude between ISEE and Ooty \citep{1990MNRAS.244..691M}.
The comparison in Paper I demonstrates the extent of overlap between a typical MWA IPS observation and ISEE and Ooty data.

Alternatively, rather than using IPS observations directly, they may be used to generate full tomographic reconstructions of the heliosphere \citep[and references therein]{1998JGR...10312049J}.
This allows IPS observations and possibly in-situ measurements \citep{2013SoPh..285..151J} over several days to be synthesised together.
If these models are derived from IPS observations at higher frequencies (generally covering sources closer to the Sun), this has the advantage that data from the recent past will allow accurate modelling of solar wind features which have now moved outwards to elongations better observed by lower-frequency instruments.

Finally there is the possibility of modelling the heliosphere from the IPS observations themselves. 
For the MWA in particular, due to the imaging technique employed, the number of IPS sources observed per day is increased by more than two orders of magnitude over a single-beam instrument.
Thus, the MWA has the potential to sample the heliosphere with unprecedented resolution.
Ideally one would start with a network of sources known to be highly compact (avoiding the need to model source structure).
It is well known that flat-spectrum sources tend to be compact.
Additionally, as shown in \papertwo, almost every source which shows a spectral peak is unresolved on IPS scales.
Indeed, as shown in \paperthree, many entirely point-like sources can be reliably identified by their spectrum (though a large population, the CSS sources, cannot).
Thus catalogues such as those of \citet{2017ApJ...836..174C} and \citet{2017MNRAS.464.1146H} provide us with a ready-made network of compact sources.
See \paperthree\ for further details including sourcecounts of these potential calibrators.

The accuracy of using our own observations depends on how the turbulence parameters (strength, velocity, power law index, axial ratio) vary as a function of sky location (i.e. on scales of a fraction of an AU).
The ability of the MWA to probe hundreds of lines of sight simultaneously, with separations between a fraction of a degree and tens of degrees makes it an excellent instrument for determining the structure function of these parameters on these large scales.
If we are able to critically sample such structure then this will allow us to correct for the effects of the varying solar wind from source to source with unprecedented accuracy.
Information on Mesoscales ($\gtrsim10^5$\,km) is provided by the variability in the IPS observables for a particular source on timescales of minutes to hours.
Such structures have been detected, at least in velocity \citep{2013SoPh..285..111H}, suggesting that there will be structure even on the smallest scales probed spatially by the MWA (1\degr$\approx$2.6$\times 10^6$\,km at 1\,AU).

\section{Observational Parameters of the MWA} \label{sec:mwa}
In this section we describe how the sensitivity of the MWA depends on the direction that the MWA is pointing (larger zenith angles mean less sensitivity) and the part of the sky towards which the MWA is pointing (higher Galactic latitudes in general means more sensitivity).
The basic details are given in \citet{2013PASA...30....7T} and other references in this section; here we give a brief summary and expand on the details which are particularly relevant to IPS observations.

Since IPS requires observations with solar elongations of a few tens of degrees, comparable with the FoV of the MWA, a critical consideration is how to deal with the presence of the Sun, a variable source which is orders of magnitude brighter than the astrophysical sources we wish to probe.

\subsection{Sensitivity of the MWA} \label{sec:mwa_sensitivity}
\begin{figure}
  \includegraphics[width=\columnwidth,trim={0 100 0 100},clip]{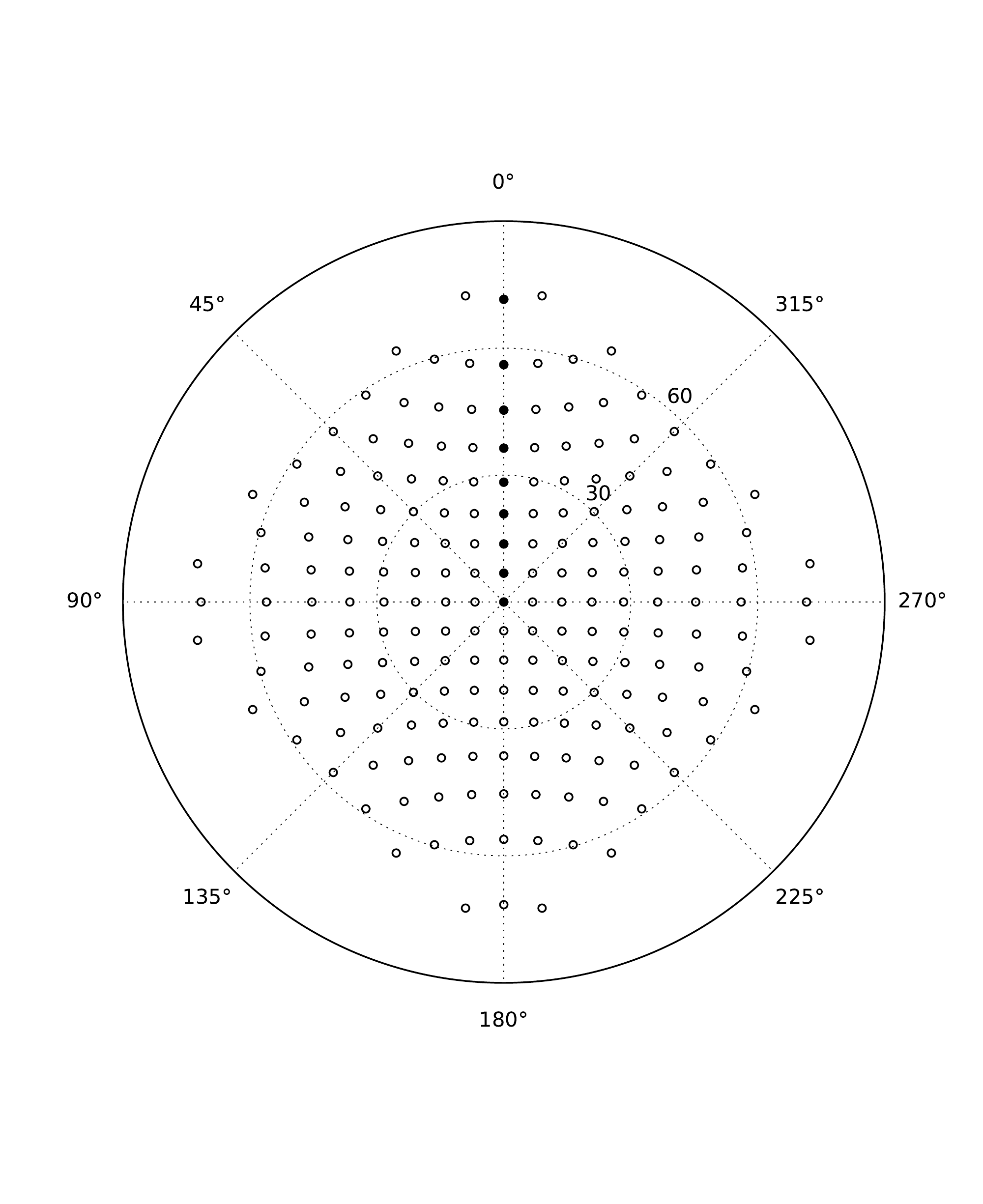}
  \caption{Figure showing azimuth and elevation of all MWA ``sweetspot'' pointings : those for which all 16 tiles are in phase for the location on the sky indicated. Those described in \tab~\ref{tab:sensitivity_el} are indicated with filled circles.}
  \label{fig:sweetspots}
\end{figure}
The MWA is not mechanically steered, instead the signals from static arrays of dipoles are combined with different delays, giving rise to a restricted set of possible pointings where all dipoles are perfectly in phase (these are shown in \fig~\ref{fig:sweetspots}).

The sensitivity of the MWA across its full FoV as a function of frequency and pointing and polarisation has been simulated with increasing sophistication and efficiency as described in a series of papers \citep{2015RaSc...50...52S,2015ITAP...63.5433S,2017PASA...34...62S}.
We draw on the most recent iteration of this modelling to give a general idea of the sensitivity of the MWA as a function of zenith angle in \tab~\ref{tab:sensitivity_el}.
The maximum sensitivity of the MWA is achieved when the MWA is pointing at the zenith.
At both frequencies, the sensitivity drops off rapidly for pointings at elevations below 45\degr.

The system noise of the MWA comes principally from diffuse Galactic emission.
This means that the system noise varies by more than a factor of two depending on where on the sky it is pointing \citep{2013PASA...30....7T}.
\citet{2015PASA...32...25W} describe how the system noise can be estimated for any MWA observation using the beam model and a simple sky model derived from the all-sky single dish survey of \citet{1982A&AS...47....1H}, scaled to the appropriate frequency.

Thus, with a model for both the beam and the system noise, and knowledge of the state of the array (i.e. how many tiles were active), we are able to predict the sensitivity at each point on the sky for any MWA observation.

\subsection{The Sun as seen by the MWA wavelengths} \label{sec:sun}
The quiet Sun at MWA wavelengths is a thermal source with a temperature $\sim10^6$\,K (the inner corona) which appears as an edge-darkened disk somewhat larger than the optical disk of the Sun \citep{1967ARA&A...5..213N,2018arXiv180804989M}.
For this wavelength and temperature the Rayleigh-Jeans approximation applies and so the Sun has a power-law spectral energy distribution with a spectral index of +2 (i.e. is it brighter at higher frequencies).
At 160 MHz the integrated quiet Sun flux density is approximately 50\,000\,Jy, many orders of magnitude brighter than the astrophysical sources we are probing.
Additionally, radio bursts, transient phenomena with complex time and frequency evolution which can last from milliseconds to days, can be many orders of magnitude brighter still than the quiet Sun \citep{1963ARA&A...1..291W}.
\citet{2017ApJ...851..151M} showcase the complex structure that can be seen on the Sun with the MWA both during a burst and a relatively quiet period.

Two observatories in Australia (as part of an international network) observe the Sun and radio bursts are automatically detected \citep{2010ApJ...710L..58L}.
These detections are then curated and archived by the Space Weather Prediction Centre (SWPC\footnote{\href{http://www.swpc.noaa.gov/products/solar-and-geophysical-event-reports}{www.swpc.noaa.gov/products/solar-and-geophysical-event-reports}}).
Thus, we can know which of our observations are affected at least by the brightest radio bursts.
See \sect~\ref{sec:space_weather} for further details.

\subsection{Observing Strategy}
\begin{table}
  \footnotesize
  \centering
  \caption{\label{tab:sensitivity_el} Sensitivity as a function of elevation all nine pointings of the MWA along the meridian from zenith to the northern horizon. Due to the fourfold symmetry of MWA tiles, all pointings towards the cardinal points will match with one of these. Other pointings match the general trend with zenith angle within $\sim$10\%. Peak gives the maximum sensitivity of the pointing compared to the centre of the zenith pointing. Area gives the number of square degrees where the sensitivity exceeds half the maximum sensitivity of the \emph{zenith} pointing.}
  \begin{tabular}{rrrrr}
    \hline
              & \multicolumn{2}{c}{80\,MHz} & \multicolumn{2}{c}{162\,MHz} \\
    Elevation & peak & area    & peak & area   \\
              &      & deg$^2$ &      & deg$^2$\\
    \hline                                                                                                                                                
    90        & 1.0  & 1330 & 1.0  & 354 \\
    83        & 0.99 & 1300 & 0.99 & 351 \\
    76        & 0.95 & 1230 & 0.95 & 342 \\
    69        & 0.88 & 1090 & 0.91 & 336 \\
    62        & 0.80 &  900 & 0.86 & 317 \\
    54        & 0.69 &  623 & 0.78 & 278 \\
    45        & 0.57 &  266 & 0.65 & 186 \\
    34        & 0.45 &  --  & 0.50 &   2 \\
    18        & 0.34 &  --  & 0.31 &  -- \\
    \hline
  \end{tabular}
\end{table}
As described in the previous section, observations of IPS at our observing frequency are best made at elongations $\sim$30\degr\ from the Sun.
Since even the quiet Sun is many orders of magnitude brighter than the astrophysical sources we wish to study, it is imperative that the Sun's flux density is strongly attenuated if we are to observe IPS without requiring unrealistically high dynamic range imaging.

First we consider placing the Sun at the centre of the FoV.
If the Sun were merely a constant source a few times brighter than the astrophysical sources that we are observing this would be a sensible strategy since placing the source to be subtracted at the centre of the beam gives the best chance of subtracting it cleanly.
Unfortunately, with the Sun orders of magnitudes brighter \emph{and} varying in time and frequency, modelling it well enough to give a dynamic range sufficient to measure IPS on 0.1\,Jy sources would in practice be impossible.
\citet{2018arXiv180201778L} have achieved a dynamic range on the Sun of 38\,000:1 only with extensive calibration on a single 40\,kHz fine channel in a single 0.5\,s integration.
Replicating this across many thousands of channels and timesteps would be untenable.
It is clearly necessary to significantly attenuate the Sun in some way.

Another possibility would be to observe while the Sun is just below the horizon.
\Tab~\ref{tab:sensitivity_el} illustrates the beam sensitivity for a selection of pointings at different zenith angles. 
Reasonable sensitivity is attainable down to approximately 45\degr\ from the horizon.
This is acceptable but means that at 162\,MHz it is not possible to observe sources at the maximum scintillation index.

However, by placing the Sun in one of the primary beam nulls, its effects are minimal, at least in those fields analysed in Papers I and II.
Moreover, the typical distance from the Sun to the centre of the field (where the sensitivity is maximum) is typically 25--30\degr, where the scintillation index is close to maximum.

The major disadvantage of this approach is that it is almost always impossible to null the Sun at two widely-spaced frequencies (since the primary beam scales approximately with frequency).
However, at lower frequencies the Sun is dimmer, whereas radio telescopes are less sensitive and even compact astrophysical sources (which tend to have flatter spectra than more extended ones) are somewhat brighter.
Thus, provided that the Sun is not too active, observations at lower frequencies with the Sun imperfect results still yield useful measurements.
In fact, in \paperone, using this approach, we detected almost as many sources in the low band as in the high band (the wider FoV in the low band being a major contributing factor).

\subsection{Imaging} \label{sec:imaging}
MWA images are typically confusion noise limited, which means that the maximum number of sources can be detected with a baseline weighting scheme close to uniform, which will maximise resolution at the expense of increased noise.
Briggs $-1$ \citep[a weighting scheme closer to uniform than natural;][]{Briggs:phdthesis} was used for the GLEAM survey \citep{2017MNRAS.464.1146H} and for Papers I\&II.

However, as a result of the work presented in \paperone, it is now clear that IPS images are very far from the confusion limit, and are limited only by the system noise.
This raises the question of whether a more natural weighting scheme might lead to greater sensitivity. \citet{2015PASA...32...25W} suggest a doubling of sensitivity going from Briggs $-1$ to Briggs $+1$.

Preliminary work confirms that the system noise in IPS observations does indeed reduce by a factor of two, and that this is a straightforward way to immediately double our sensitivity.
There are a few costs to this approach, however.
Most obviously, moving to a more natural weighting scheme must mean that the interferometric resolution is decreased.
This will not greatly affect the astrometric accuracy, so for an isolated IPS source it should not be any more difficult to match it with a known source 
(astrometric accuracy scales linearly with both S/N and resolution; see \citealp{1988ApJ...330..809R}).
The reduced resolution will, however increase the chance that two nearby IPS sources will be confused (manifesting as a reduced scintillation index; see \sect~\ref{sec:source_size}).
Furthermore, Briggs $+1$ MWA images have more pronounced sidelobes than more uniformly-weighted MWA images.
This is due to the core of the MWA being extremely dense relative to the outer antennas.
This leads to areas of very high density in the UV plane.
Methods have been developed to distinguish between true scintillating sources and the sidelobes of scintillating sources (see \paperone), however some increase in noise is unavoidable. 

\section{The MWA Phase I Compact Source Survey} \label{sec:survey}
Now that we have laid out the necessary background in \sect~\ref{sec:ips} and described the pertinent characteristics of the MWA in \sect~\ref{sec:mwa} we are able to fully describe the coverage and sensitivity of our survey based on the observations we have made.
\subsection{General Observing Parameters} \label{sec:observations}
\begin{figure*}
  \includegraphics[width=\textwidth]{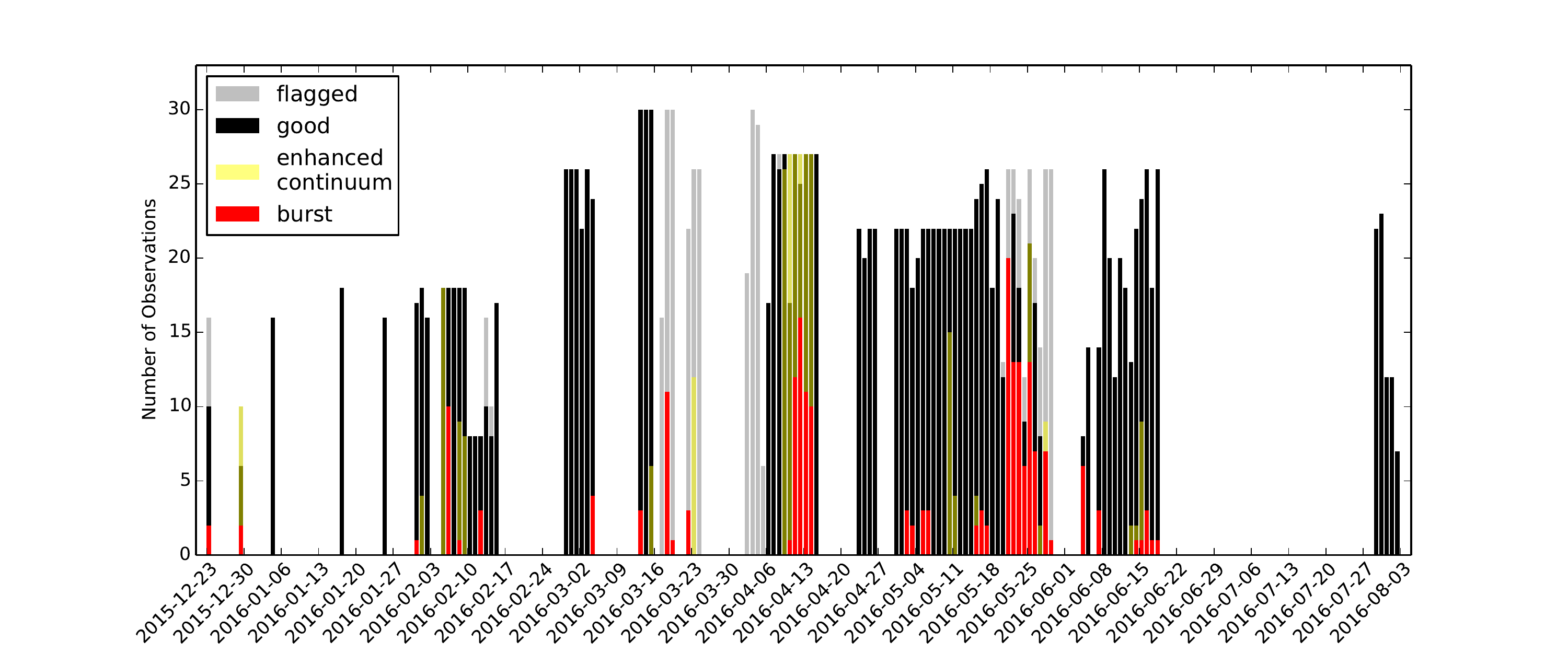}\\
  \includegraphics[width=\textwidth]{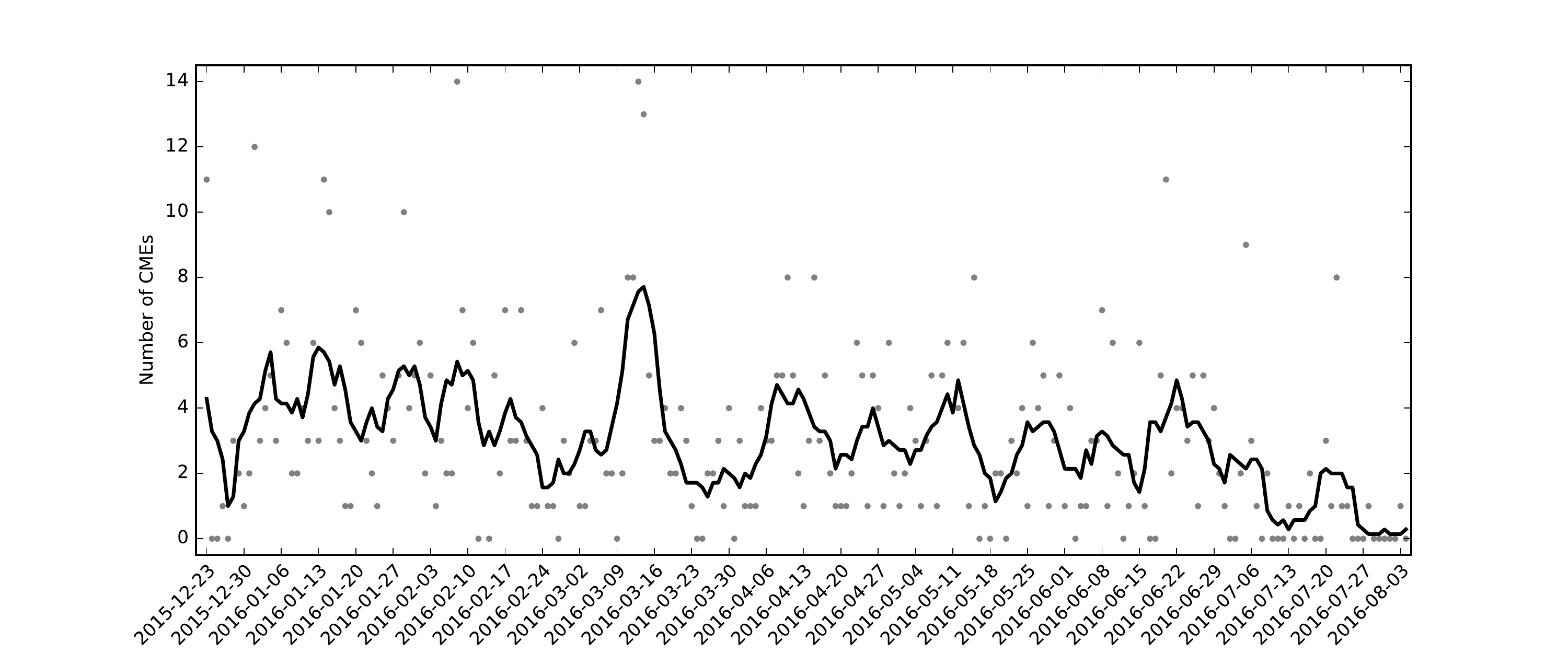}\\
  \includegraphics[width=\textwidth]{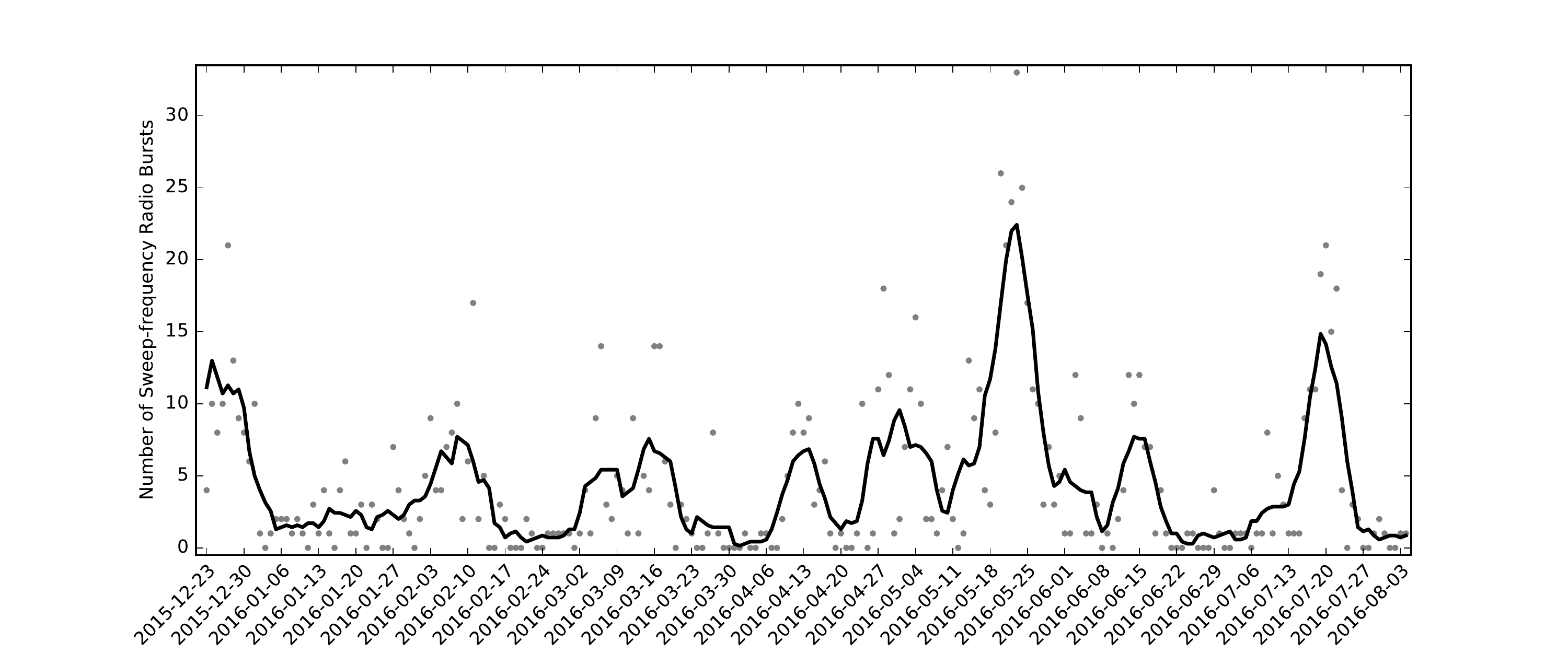}\\
  \caption{Top panel: Number of observations on each date (UTC+08:00). Greyed observations are not included in further analyses; other colored observations are coincident with solar radio activity. Middle panel: CMEs per day detected by CACTUS. Bottom panel: Sweep Frequency Radio Bursts per day. Black dots show daily totals, grey lines show moving average}
  \label{fig:date_histogram}
\end{figure*}
Observations commenced on 2015-12-23 and finished on 2016-08-02 when the MWA was shut down for reconfiguration into a new Phase II array (Wayth et al. in prep.).
Initially a set of observations were made once per week, but this was soon increased to one set of observations per day when the array was available.
On occasion, the array was not available for some or all of the day, either due to clashes with other observations or hardware failures.
Most notably the array suffered lightning damage in mid-March which could not be repaired for several weeks.
Overall, usable observations were made on 91/224 local (UTC+08:00) calendar dates (see \fig~\ref{fig:date_histogram}).

Observations were scheduled in pairs of consecutive 5-minute observations.
The Western limb is surveyed over the course of the morning and Eastern limb in the afternoon.

\subsection{Space Weather Context} \label{sec:space_weather}
\begin{figure}
  \includegraphics[width=\columnwidth]{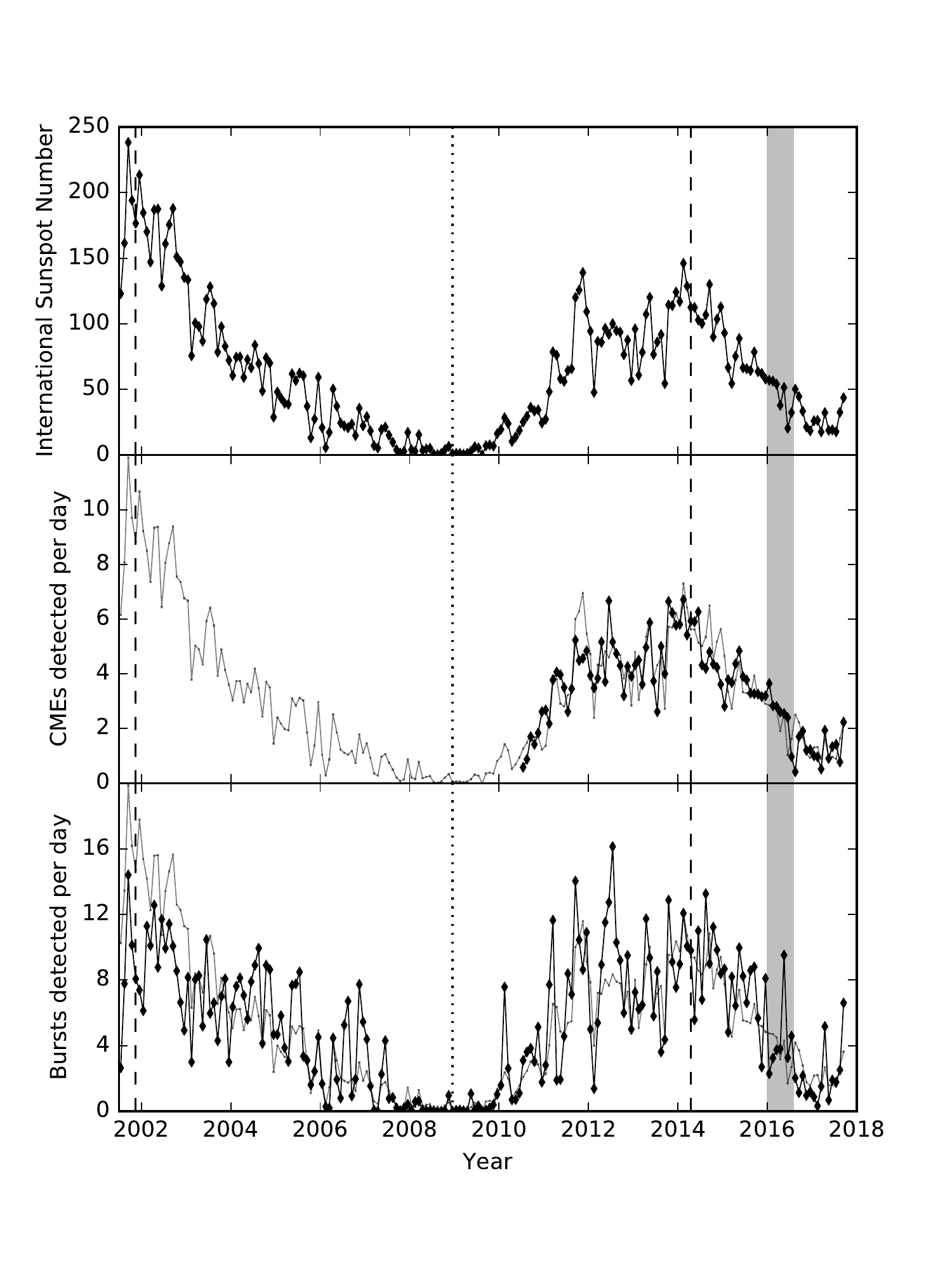}\,
  \caption{Top panel: International Sunspot Number (ISSN); Middle Panel: black line: CMEs per day detected by CACTUS; grey line: scaled ISSN. Bottom panel: black line: Sweep Frequency Radio Bursts per day; grey line: scaled ISSN. Dotted and dashed lines indicated solar minimum and maximum respectively. All quantities averaged by calendar month.}
  \label{fig:space_weather_context}
\end{figure}
\fig~\ref{fig:space_weather_context} shows the International Sunspot Number \citep{sidc}, the number of CMEs detected by Computer Aided CME Tracking software \citep[CACTus][]{2004A&A...425.1097R} using data from space-based imagers \citep{1995SoPh..162..357B,2008SSRv..136...67H}, and the Radio burst events recorded in the Space Weather Prediction Centre archive\footnote{\href{http://www.swpc.noaa.gov/products/solar-and-geophysical-event-reports}{www.swpc.noaa.gov/products/solar-and-geophysical-event-reports}}.
Note that the latter two are only illustrative since they use the raw numbers from their respective archives: no account is taken of instrument availability etc.

These data show that this survey took place when Solar Cycle 24 was in its declining phase, having peaked in April 2014.
Nonetheless, there are still a significant number of CMEs (meaning that there will be inhomogeneities in the solar wind) and large numbers of solar radio bursts, which have the potential to interfere with our observations.
The number of detected CMEs and solar radio bursts are shown in more detail in \fig~\ref{fig:date_histogram}.
We find that 205 (10\%) of our observations are concurrent with a burst event; almost all of which are of type III.
The majority are actually aggregated into so-called ``Type VI'' bursts which are defined as a ``Series of Type III bursts over a period of 10 minutes or more, with no period longer than 30 minutes without activity''. So overlap with a Type VI event does not necessarily mean that a burst was coincident with our observation.
Moreover, the duration of a Type III burst is just a few seconds, so only a small fraction of an observation would need to be discarded for each one.
%

\subsection{Sky Coverage}
\label{sec:coverage}
\begin{figure*}
  \includegraphics[width=\textwidth]{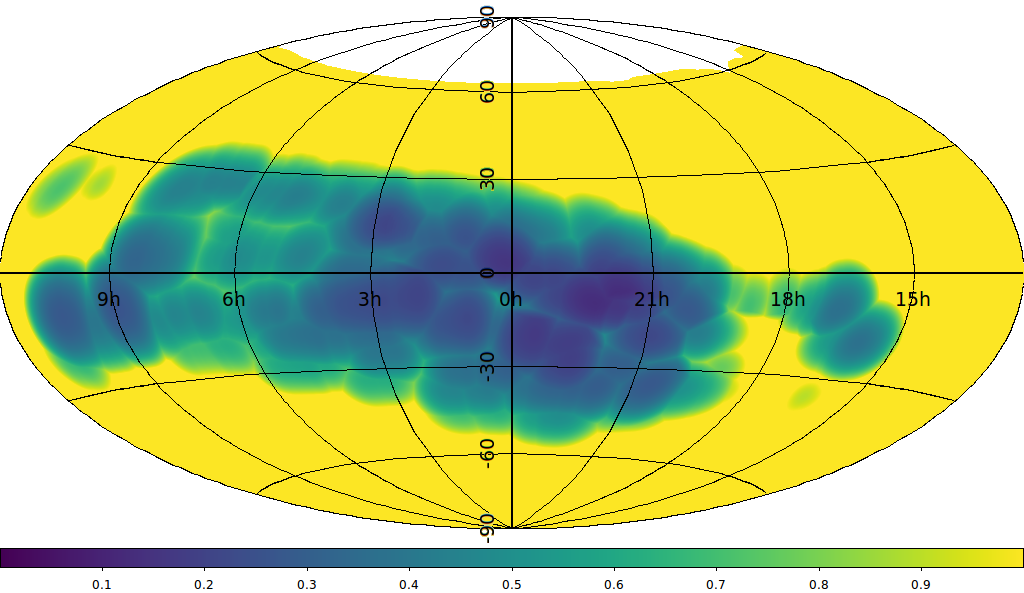}
  \includegraphics[width=\textwidth]{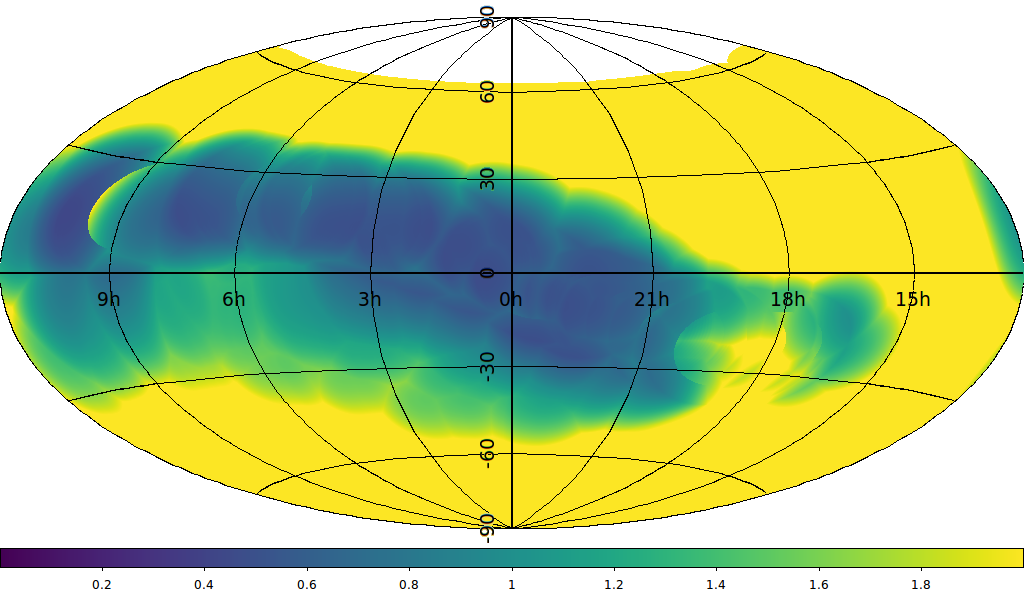}
  \caption{Maximum sensitivity achieved for each point on the celestial sphere, regardless of which observation it is in for the high-band (162\,MHz, top) and low-band (80\,MHz, bottom). Uncoloured areas to the North are below the horizon in all observations.}
  \label{fig:m_survey}
\end{figure*}
\begin{table}
  \footnotesize
  \centering
  \caption{\label{tab:coverage} Derived from \fig~\ref{fig:m_survey}, sky coverage over which we obtain various sensitivities for both the high-band (162\,MHz) and low-band (80\,MHz).}
  \begin{tabular}{rrr}
    \hline
    Sensitivity & area 80\,MHz & area 162\,MHz \\
    mJy         & deg$^2$      & deg$^2$       \\
                                                \hline                                                                                                                                   
    200         & --           &   680         \\
    400         & --           &  6500         \\
    1000        & 9700         & 17000         \\
    2000        & 20000        & 23000         \\
    4000        & 26000        & 27000         \\
    \hline
  \end{tabular}
\end{table}
We now use the methodology laid out in \sect~\ref{sec:mwa_sensitivity} to estimate the rms noise in a single 0.5\,s integration for each point on the sky for each observation, accounting for any antennas that are flagged.
From this quantity (denoted $\mu$ in \paperone) it is straightforward to calculate the 5-$\sigma$ limit to which we can detect scintillating flux density (via \paperone~\eqn~5; see also \paperthree~\eqn~1).
We then calculate the \emph{compact} flux density to which each point on the sky is sensitive using \eqn~\ref{eqn:mano} with an upper limit on the scintillation index of 0.8 and assuming solar minimum.
We further assume that in areas of sky where the scintillation would be in the strong regime ($m$>1.5 as given by \eqn~\ref{eqn:mano}), no detection is possible.
To summarise, we model the observed scintillation $m_\mathit{obs}$ to be
\begin{equation}
	m_\mathit{obs} = \left\{ \begin{array}{ll} 
    m, & m<0.8 \\
    0.8, & 0.8\leq m<1.5 \\
    0, & \hbox{otherwise,} \\
    \end{array} \right. 
	\label{eqn:m_model}
\end{equation}
where $m$ is computed via \eqn~\ref{eqn:mano}, with $\rho=1.5$, $b=1.6$.
When we compare our 5-$\sigma$ limit to that determined for the field analysed in \papertwo\ (which is part of this survey) we find that the \papertwo\ sensitivity was lower by a factor of 1.19 than that expected. 
Therefore we scale all of our sensitivity maps by this factor.
The result of this process is a set of 2095 all-sky maps, one for each observation.
We exclude those maps where the array was not fully functional as described in \sect~\ref{sec:observations} and we exclude any observations where the attenuation of the sun was less than a factor of 100.
%
The remaining 1702 maps are summarised in \fig~\ref{fig:m_survey}, showing the maximum sensitivity for each point on the celestial sphere, regardless of what observation it lies in.
\Tab~\ref{tab:coverage} shows the area over which we obtain sensitivities ranging from 200\,mJy to 4\,Jy.
Note that while these maps should give a reasonable estimate of our sensitivity, we err on the side of pessimism for the following reasons:
1) we assume the use of the uniform baseline weighting scheme used for Papers I\&II since the potential for increasing our sensitivity with a more natural weighting scheme (see \sect~\ref{sec:imaging}) is untested;
2) we assume the largest filling factor (solar minimum) of the fast (polar) solar wind, which is much less dense, (see \sect~\ref{sec:ips});
3) rather than allowing scintillation indices to approach or exceed 1, we assume a maximum scintillation index of 0.8 (\sect~\ref{sec:optimal_detection});
4) we assume that sources cannot be detected once in the strong regime;
5) observations are excluded if the Sun's primary beam attenuation factor is less than 100; and
6) we assume no upscattering, whereas if a source is observed multiple times, it will most likely be detected below the mean detection limit (\sect~\ref{sec:optimal_detection}).

Note that the lack of sensitivity around RA~18\,h is not due to a lack of observations but a lack of sensitivity of the array due to the presence of the Galactic Plane close to the Galactic Centre in the FoV; a less pronounced lack of sensitivity can also be seen around RA~06\,h where our observations cross the Galactic Plane closer to the Galactic Anti-centre (see \sect~\ref{sec:mwa_sensitivity}).

The survey area covers a range of Galactic latitudes, crossing the Galactic Plane close to the Galactic Centre and Anti-centre.
Coverage of the Southern Galactic Pole is particularly good, with excellent overlap with well-studied fields such as the MWA EoR-0 field \citep[e.g.][]{2016MNRAS.458.1057O}, GAMA \citep{2009A&G....50e..12D}, SDSS BOSS \citep{2013AJ....145...10D}, and Stripe 82\footnote{\href{http://cas.sdss.org/stripe82}{cas.sdss.org/stripe82}}.

\subsection{Expected Source Counts}
Using the coverage maps shown in \fig~\ref{fig:m_survey} and the source counts presented in \paperthree, it is possible to predict, with some extrapolation down to lower flux densities, the number of sources that we will detect across the full survey.
Here we restrict ourselves to the high band, where we have measured source counts, and to areas of sky where our detection limit is 1\,Jy or lower.
We expect that the total number of compact detections will be 11\,000 sources.
Of these, 4\,500 will be point sources, of which 1\,100 will be peaked, 2\,100 will be CSS, and 1\,200 will be flat-spectrum.
This is an increase of over a factor of 50 compared to the single field studied in Paper III and will allow sub-populations to be studied in much greater detail.

These counts do not include pulsars, which will be much more prevalent at lower Galactic latitudes than in the high Galactic latitude field from which these source counts were derived.

\subsection{Computational Requirements}
The steps required to process $\sim$100\,GB of raw data from the MWA archive (the typical amount for an individual 5-minute observation) and produce a catalogue of compact sources are discussed in \paperone, Appendix A1.
The sizes of various data products are summarised in \tab~\ref{tab:data_size}.
The total computing time needed to process the entire survey is similar to other large MWA projects, around 36\,hours using the full capacity of a $\sim$500-node supercomputer such as Galaxy at the Pawsey Supercomputing Centre\footnote{\href{http://www.pawsey.org.au/our-systems/galaxy-technical-specifications}{www.pawsey.org.au/our-systems/galaxy-technical-specifications}}.
\begin{table}
  \footnotesize
  \centering
  \caption{\label{tab:data_size} Size of data products. The raw visibilities are currently stored in the MWA data archive. Those in bold are derived data products which will ultimately be made openly available}
  \begin{tabular}{lrr}
    \hline
    Data Product                                & Size per       & Size for \\
                                                & observation    & full survey \\
    \hline                                                                                                                                                    
    Raw visibilities                            & 111\,GiB       & 224\,TiB       \\
    Calibrated measurement sets                 & 66\,GiB        & 135\,TiB       \\
    Raw FITS images                             & 88\,GiB        & 180\,TiB       \\
    \textbf{Archived image data}                & 18\,GiB        & 38\,TiB        \\
    \textbf{Variability images}                  & 39\,MiB        & 78\,GiB        \\
    \textbf{solar wind parameters}              & $\sim$100\,KiB & $\sim$200\,MiB \\
    \textbf{Astrophysical Catalogue}            & --             & $\sim$1\,MiB   \\
    \hline
  \end{tabular}
\end{table}

\section{Future Work} \label{sec:future}
In \sect~\ref{sec:ips} we drew on the literature and our own MWA observations to describe the IPS observables and the astrophysical parameters of sources that we can glean from them.
In \sect~\ref{sec:mwa} we described the relevant parameters of the MWA instrument.
This allows us to predict the likely sensitivity of our Phase I survey as described in \sect~\ref{sec:survey}.

As we analyse the data from our Phase I survey we should be able to resolve the remaining uncertainties in the optimum surveying approach, namely: 
1) the trade-off between getting close to the Sun to maximise scintillation index vs dealing with the strong scintillation regime and the Sun being imperfectly nulled;
2) the optimal image weighting scheme to maximise sensitivity while avoiding sidelobe confusion. 
Additionally we will measure the structure function of the solar wind on scales $\sim$1--10\degr\ as seen from Earth (i.e. small fractions of an AU).
This will allow us to determine the error on quantities derived from our IPS observations (e.g. source size) \emph{after} we have used neighbouring sources to account for stochastic changes in the solar wind.

\subsection{Phase II MWA}
The original MWA Phase I configuration used for the survey described in \sect~\ref{sec:survey} was designed to satisfy the requirements of the four main science themes \citep{2013PASA...30...31B} including Epoch of Reionisation studies, which requires high sensitivity on very large spatial scales.
As a result, 65\% of MWA Phase I baselines are $<750$\,m in length (corresponding to roughly 10\arcmin\ resolution at 150\,MHz)
From mid-2016, a further 128 tiles were added to the MWA (Wayth et al. in prep.): some to provide baselines up to 5\,km, others to provide further short baselines for EoR studies.
It is envisaged that the MWA will switch between just two configurations: a short baseline configuration and a long baseline configuration.
Thus any future MWA IPS survey will use one of these two configurations.

The new long-baseline configuration is particularly well-suited to IPS observations since it provides extremely uniform coverage of the UV plane.
This means that IPS observations with the long baseline Phase II MWA will combine the sensitivity of naturally-weighted Phase I images with the exceptional sidelobe suppression of uniformly-weighted Phase I images.

The only downsides are that in order to use the very longest baselines it would be necessary to image with higher resolution: increasing the computational requirements and increasing the magnitude of ionospheric effects. 
However even if baselines $>3000$\,m were excluded (which would reduce the image resolution requirements to those used in Papers I and II) only 15\% of baselines would be lost.

Excluding these long baselines does not exclude all baselines to the outer tiles, so the footprint of the array on the ionosphere is significantly larger, increasing ionospheric scintillation and direction-dependent calibration errors.
However it was shown in \paperone\ that even in very adverse ionospheric conditions the effects on our IPS analysis are minimal.

\subsection{Possible future upgrades to the MWA}
An obvious next upgrade to the MWA would be to correlate all 256 existing tiles simultaneously.
This would result in 3$\times$ as many baselines in the range that interests us (350$\lambda$--1\,500$\lambda$), though the dense concentrations of tiles at the core will reintroduce the sidelobes present in the Phase I configuration for naturally-weighted images.

Increasing the bandwidth of the MWA may increase the IPS sensitivity slightly, however the difficulties of keeping the Sun nulled across a wide bandwidth mean that this is unlikely to increase the sensitivity drastically.
Perhaps more useful would be an upgrade to the beamformers to allow multiple beams to be observed across the sky, increasing the instantaneous FoV.

If the correlator were upgraded to allow a shorter integration time, this would allow observations further into the strong regime (and improved resolution to detect fine structure for the very brightest sources). 
As computing power gets cheaper, observing at higher time resolution, and imaging more spectral channels will become more computationally feasible.

\subsection{SKA-low}
SKA-low is expected to be more than 100$\times$ more sensitive than the 128-tile MWA, which would allow scintillating sources to be detected as low as $\sim$0.2\,mJy.
The technical requirements which would allow the SKA-low to carry out image-based IPS observations are relatively modest: it is required only that the SKA be able to form images with a cadence of 0.5\,s or less, and that it can observe within a few tens of degrees of the Sun.
However the value of such observations would be very great: allowing spatial scales well beyond the interferometric resolution of SKA-low to be probed.
Papers II to IV of this series showcase some of the  extragalactic science which this enables.
An additional important science case, using IPS to efficiently discover pulsars, is discussed in \sect~5.4 of \papertwo.

In contrast to our MWA observations, it cannot be taken for granted that the SKA-low will not be confusion limited.
Thus it is useful for us to consider the resolution that would be required for classical confusion to be reduced to the same 0.2\,mJy level as the system sensitivity. 
This requires compact source counts several orders of magnitude in flux density below those we have measured in \paperthree, and since no other information on compact sources exist at mJy levels at this frequency, large uncertainties are unavoidable.
\begin{figure}
  \includegraphics[width=\columnwidth]{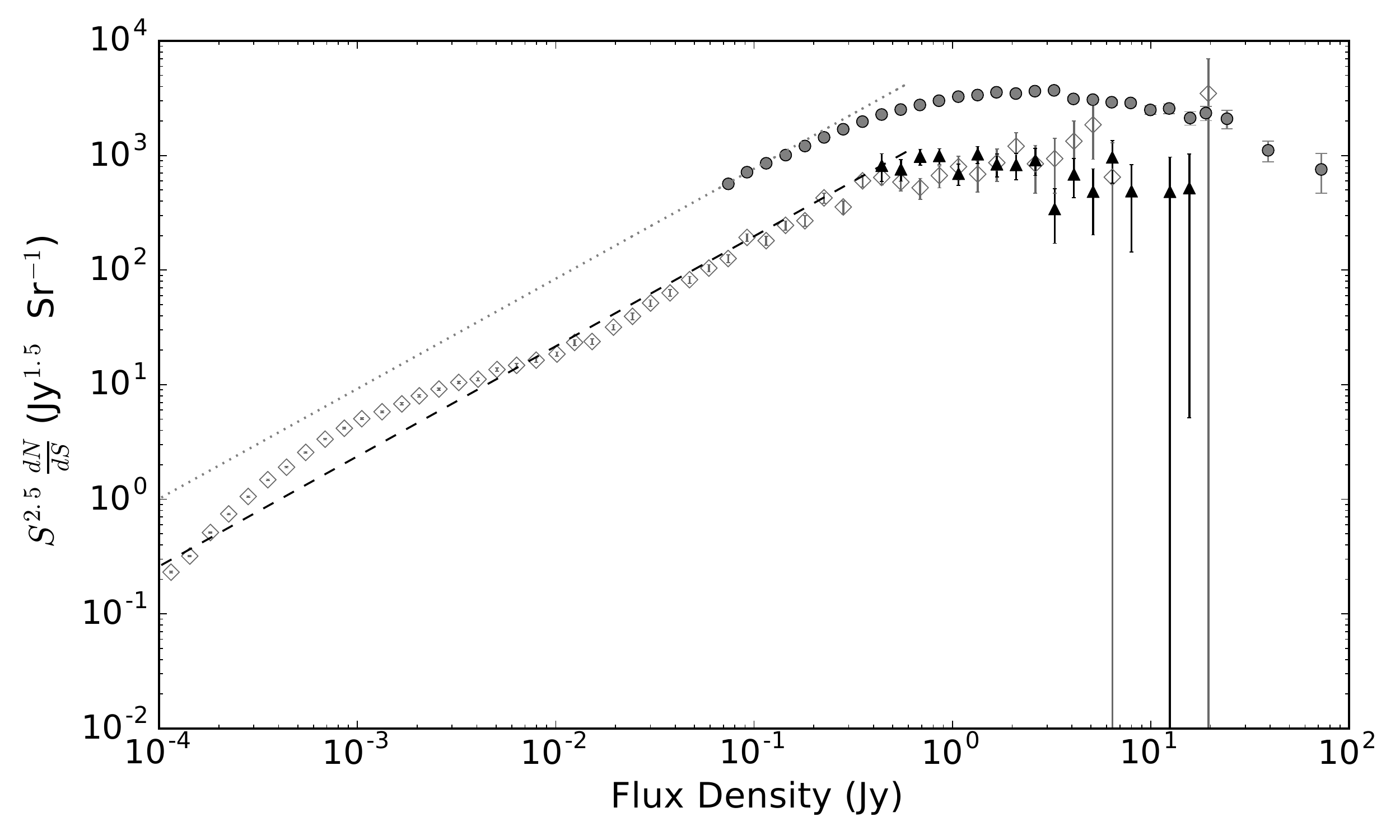}\,
  \caption{Euclidean-weighted differential source counts. Grey circles: GLEAM source counts (Franzen et al. in prep. as presented in \paperthree); dotted line source counts at 151 MHz as measured by \citet{2016MNRAS.459.3314F}; black triangles: source counts of IPS sources as presented in \paperthree; dashed line: dotted line scaled to match lowest black triangle; diamonds: 151\,MHz IPS source counts predicted by simulation of \citet{2008MNRAS.388.1335W}.}
  \label{fig:s3_counts}
\end{figure}

In the absence of direct measurements, we calculate the source counts in the relevant flux density range using two different approaches, both of which are illustrated in \fig~\ref{fig:s3_counts}. The simplest approach assumes that the compact source counts have a power law index, $\gamma$, exactly matching the standard source counts at 150\,MHz \citep{2016MNRAS.459.3314F}, but with the scaling factor, $k$, reduced by a factor of 2.5 to match the lowest flux density counts we were able to measure (i.e. $k=1800$, $\gamma=1.5$).

The other approach we take is to measure the source counts predicted by the S$^3$ simulation of \citet{2008MNRAS.388.1335W}\footnote{\href{http://s-cubed.physics.ox.ac.uk/}{s-cubed.physics.ox.ac.uk}}.
We calculate these simulated counts exactly as described in \sect~3.1 of \paperthree\ except that a resolution is 100\arcsec\ is assumed.
As shown in \fig~\ref{fig:s3_counts}, these two approaches agree almost perfectly below 200\,mJy, apart from around 1\,mJy, where the simulation predicts a population of starforming disks which are sufficiently distant/compact to show IPS.

A full analysis would also take into account the unique way in which noise and confusion manifest when detecting sources via their variability.
As described in \paperone\ (see also \sects~\ref{sec:optimal_detection}\&\ref{sec:coverage} above), we search for scintillating sources not in a standard interferometry image but in a ``variability image'' which has Gaussian noise statistics, but where the value of a pixel corresponding to a scintillating source does not give the variable flux density $\Delta S$, but a related value, which close to the detection limit is proportional to $\sqrt{\Delta S}$.
The sources which contribute to confusion are very much in this regime which has the effect of flattening the source counts.
Additionally, unless extremely close to each other, the observed flux density will be the flux densities of the individual sources combined in quadrature (see \tab~\ref{tab:source_nsi}) which will also have the effect of reducing classical confusion.

Classical confusion is usually considered to be significant when $s/b$, the number of sources per beam, exceeds $1/50$ -- $1/10$ \citep[e.g][]{2001AJ....121.1207H}.
Due to the effects above, we adopt the limit of $s/b$  $\le 1/10$. 

The SKA Baseline design envisages a compact core with a resolution of 10\arcmin\ surrounded by 3 logarithmic spirals with stations out to a radius of 90\,km\footnote{SKA-TEL-SKO-DD-001 \url{http://www.skatelescope.org/wp-content/uploads/2012/07/SKA-TEL-SKO-DD-001-1_BaselineDesign1.pdf}}.
A more recent proposed configuration\footnote{SKA-SCI-LOW-001 \url{http://indico.skatelescope.org/event/384/attachments/3008/3961/SKA1_Low_Configuration_V4a.pdf}} has an even more compact core and a maximum radius of 35\,km.

With a resolution of 10\arcmin\ the classical confusion limit would be reached at approximately 20\,mJy\footnote{NB all of these limits are computed using the \citet{2001AJ....121.1207H} definition of beamsize which is the 1-$\sigma$ radius of an elliptical Gaussian}.
Clearly, much longer baselines will be required to achieve reasonable sensitivity.
At 300\arcsec\ resolution (optimum sensitivity), the confusion limit is 1.3\,mJy. 
This drops to 0.08\,mJy with 150\arcsec\ resolution.
At this resolution the SKA-low configuration referenced above can achieve a sensitivity of approximately 0.5$\times$ that of natural weighting with a sidelobe level is $\sim$1\%.
With $\gamma=1.5$ there is approximately one source with a S/N$\ge$100 per 100 resolution units.
This suggests system noise, classical confusion, and sidelobe confusion are of a similar order of magnitude.
However the main point is that imaging the full 27 square degree FoV of the SKA-low with 150\arcsec\ resolution is not an onerous task; on the contrary, it requires only moderately longer baselines than the current MWA, and produces images with 1/10th of the pixels of a standard MWA image. 

Since a 4\,km baseline (corresponding to a resolution of 100\arcsec\ at 150\,MHz) is very small compared to the spatial scales of the turbulence responsible for IPS, we should not need to consider the loss of spatial coherence across the array due to IPS. 
However, if the SKA data processor allows it, the very longest SKA baselines might be used commensally for multi-station IPS, using beamforming on each of the brighter IPS sources.
A 64\,km baseline is sufficient to measure the speed \citep{1967Natur.213..343D,2016ApJ...828L...7F}, however a baseline of 180\,km or more would be necessary to resolve the high- and low-speed streams \citep{1996Natur.379..429G}.

\section{Conclusions} \label{sec:conclusions}
Over this series of publications we have demonstrated that with a single 5-minute MWA observation we can characterise the sub-arcsecond properties of over 1000 sources.
Having identified those sources which contain compact structure, we have explored their radio and multi-wavelength properties.
The aim of this paper has been to gather together the necessary background information on the theory and practice of IPS, and the observational constraints of the MWA, to permit the analysis of a full survey taken over 8 months and covering $\sim$15\,000 square degrees.

The main purpose of this survey will be to identify compact sources, both extra-galactic and pulsars, for further study (see Papers III and IV for further details). 
The scientific productivity of this endeavour has already been verified in Papers II, III and IV, and most of the lines of enquiry being followed would benefit from the 50$\times$ increase in sample size afforded by the full survey.
However, as we have shown in \sect~\ref{sec:source_size} there are many techniques that may be applied to determine a more detailed morphology for each source.
Furthermore we should not discount the possibility of unexpected discoveries as we probe into unexplored parameter space.

This survey was scheduled manually and in an ad hoc manner, working around other MWA projects.
At the time, the accuracy of MWA beam models had not been verified.
However the retrospective analysis in \sect~\ref{sec:survey} shows that the survey strategy was a good one, and the resulting observations have excellent coverage across a wide range of Galactic latitudes, with good overlap with well-studied fields at high-Galactic latitude, within which we can detect compact structure down to a flux density limit $\sim$400\,mJy or lower.

In \sect~\ref{sec:future} we show that with future observations we will probe considerably deeper, and that with instruments such as SKA-low there is no reason why this technique cannot be used to probe compact sources down to sub-mJy levels.

\section*{Acknowledgements}
Parts of this research were conducted by the Australian Research Council Centre of Excellence for All-sky Astrophysics (CAASTRO), through project number CE110001020.

\bibliographystyle{hapj}
\bibliography{refs} 

\label{lastpage}
\end{document}